\documentclass[aps,pra,twocolumn,reprint,floatfix, superscriptaddress,longbibliography]{revtex4-2}

\usepackage{subfigure}
\usepackage{psfrag,graphicx}
\usepackage{pict2e}
\usepackage{dcolumn}
\usepackage{amsmath,amssymb,braket}
\usepackage{bm}
\usepackage{color}
\usepackage{latexsym}
\usepackage{epstopdf}
\usepackage{color}
\usepackage{comment}
\usepackage[english]{babel}
\UseRawInputEncoding
\usepackage{amsfonts}
\usepackage{bm}
\usepackage{natbib}
\usepackage{bbold}

\usepackage{enumerate}
\definecolor{myblue}{rgb}{0.2,0.2,0.8}
\definecolor{myzard}{cmyk}{0,0,0.05,0}
\definecolor{mywhite}{rgb}{1,1,1}
\definecolor{mywhite}{rgb}{1,1,1}
\definecolor{myred}{rgb}{1,0.,0.3}
\usepackage[colorlinks=true,citecolor=myblue,linkcolor=myred]{hyperref}

\definecolor{darkgreen}{rgb}{0.0, 0.4, 0.26}

\definecolor{mygrey}{gray}{0.35}
\definecolor{myblue}{rgb}{0.2,0.2,0.8}
\definecolor{myzard}{cmyk}{0,0,0.05,0}
\definecolor{mywhite}{rgb}{1,1,1}
\definecolor{mywhite}{rgb}{1,1,1}
\definecolor{myred}{rgb}{1,0.,0.3}

\def\be{\begin{equation}}
\def\ee{\end{equation}}
\def\ba{\begin{align}}
\def\enda{\end{align}}
\def\bi{\begin{itemize}}
\def\ei{\end{itemize}}

\def\beq{\begin{equation}}
\def\beq{\begin{equation}}
\def\eeq{\end{equation}}

\def\kk{{\textbf k}}

\def\rr{{\textbf r}}

\newcommand{\carlos}[1]{{\color{black} #1}}

\begin{document}

\title{Topological multi-mode waveguide QED}

\author{C.~Vega}
\affiliation{Institute of Fundamental Physics IFF-CSIC, Calle Serrano 113b, 28006 Madrid, Spain.}

\author{D.~Porras}
\affiliation{Institute of Fundamental Physics IFF-CSIC, Calle Serrano 113b, 28006 Madrid, Spain.}

\author{A.~Gonz\'{a}lez-Tudela}
\affiliation{Institute of Fundamental Physics IFF-CSIC, Calle Serrano 113b, 28006 Madrid, Spain.}

\begin{abstract}
Topological insulators feature a number of topologically-protected boundary modes linked to the value of their bulk invariant. While in one-dimensional systems the boundary modes are zero-dimensional and localized, in two-dimensional topological insulators the boundary modes are chiral, one-dimensional propagating modes along the edges of the system. Thus, topological photonic insulators with large Chern numbers naturally display a topologically-protected multi-mode waveguide at their edges. Here, we show how to take advantage of these topologically-protected propagating modes by interfacing them with quantum emitters. In particular, using a Harper-Hofstadter lattice, we find situations in which the emitters feature quasi-quantized decay rates due to the increasing number of edge modes, and where their spontaneous emission spatially separates in different modes. We also show how using a single $\pi$-pulse the combination of such spatial separation and the interacting character of the emitters leads to the formation of a single-photon time-bin entangled state with no classical analogue, which we characterize computing its entanglement entropy. Finally, we also show how the emitters can selectively interact with the different channels using non-local light-matter couplings as the ones that can be obtained with giant atoms. Such capabilities pave the way for generating quantum gates among topologically-protected photons as well as generating more complex entangled states of light in topological channels.
\end{abstract}
\maketitle

\section{Introduction}

Topological photonics~\cite{ozawa19a,rider2019,Smirnova2020a,Price2022} is a burgeoning field aiming at exporting topological concepts into photonics to bring novel, and more robust, ways of controlling the properties of light. At the classical level, one of the original motivations of the field was to exploit the chiral, edge modes appearing in two-dimensional topological insulators to obtain unidirectional and robust-to-disorder photon flows~\cite{haldane08a,Raghu2008}. In their simplest instance, that is, when only a single edge mode appears, it was soon realized in several platforms~\cite{wang09a,rechtsman13a,hafezi13b}. Remarkably, these realizations opened up applications beyond their initial motivation, such as the design of ``topological lasers"\cite{Bahari2017,Bandres2018,Klembt2018,Amelio2020} and chiral light-matter couplings~\cite{Barik2018,Barik2020,Hallett2022,Mehrabad2020,Owens2021}, or the generation of gaussian quantum correlations with parametric drivings~\cite{mittal16a,mittal18a,Mittal2021,Blanco-Redondo2018,Wang2019b,Blanco-Redondo2020,Doyle2022}. Besides, they have been proposed as robust \emph{quantum buses} for quantum state transfer~\cite{Yao2013,Lemonde2019,Cao2021,Dlaska2017}, without the exponential time dependence on their distance of their zero-dimensional counterparts~\cite{Almeida2016,Mei2018,Lang2017,Longhi2019,Tran2020c,Wang2021b,Han2021}. 

The more complex scenario when the two-dimensional system features several edge modes~\cite{Skirlo2014,Skirlo2015} opens up new opportunities, e.g., to increase quantum communication capacity using several topologically-protected channels. Remarkably, this situation has been much more scarcely explored in the literature, likely because it is still unclear how to profit from these extra modes. Coupling quantum emitters to such structures provide a natural route to harness these additional degrees of freedom. On the one hand, local light-matter coupling enables the coupling to all photonic modes at the emitter frequencies, thus being able to interact with several modes simultaneously. This has already been shown to lead to unconventional emitter-emitter interactions when coupled to the bulk modes of one-\cite{Bello2019a,Leonforte2020b,Vega2021a}, two-\cite{Leonforte2020b,DeBernardis2021}, and three-\cite{Garcia-Elcano2020,Garcia-Elcano2021} dimensional topological insulators. Besides, the strongly interacting character of the emitters can induce (non-gaussian) quantum correlations beyond the ones that can be obtained with parametric drivings~\cite{mittal16a,mittal18a,Mittal2021,Blanco-Redondo2018,Wang2019b,Blanco-Redondo2020,Doyle2022}, opening a path to observe exotic quantum many-body states~\cite{Bello2022,Grusdt2013,Grusdt2014,Maghrebi2015,Schine2016,Clark2020}. All these reasons are motivating the development of such topological light-matter interfaces in various platforms, ranging from superconducting qubits coupled to microwave resonators~\cite{Kim2020b,Owens2021} to solid-state emitters coupled to topological photonic crystals~\cite{Barik2018,Barik2020,Hallett2022,Mehrabad2020}.

\begin{figure}[tb]
\includegraphics[width=0.9\columnwidth]{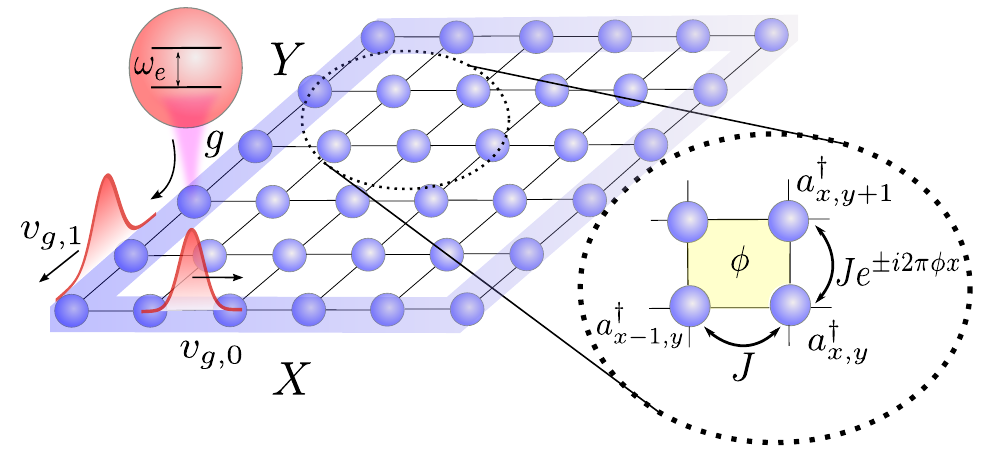}
\caption{Topological light-matter interface: Blue globes represent lattice sites, with annihilation (creation) operators $a_{x,y}^{(\dagger)}$. The yellow-shaded region denotes a single lattice plaquette with flux $\phi$. We also represent the spontaneous emission of an emitter (in red) coupled to the the lattice edge, which radiates through all edge modes ($2$ in the figure), that propagate at different group velocities (denoted by $v_{g,0}$ and $v_{g,1}$), causing a spatial separation of the emitted pulse.}
\label{fig:1}
\end{figure}

In this work, we develop a theory for the topological multi-mode waveguide QED scenario that appears when quantum emitters couple to the edges of topological photonic insulators with large Chern numbers. The physics emerging from this scenario is very different from the case of emitters coupled to one-dimensional topological insulators~\cite{Bello2019a,Leonforte2020b,Vega2021a}, where the localized nature of the boundary modes leads generally to coherent emitter dynamics/interactions rather than irreversible or collective decay dynamics. Using that theory, in Section~\ref{sec:model} we unveil, for the first time, the entanglement structure of the spontaneously emitted photons in such topological multi-mode waveguides, and show one can obtain almost maximally entangled $W$-type states between the edge channels. Besides, combining several $\pi$-pulse~\cite{Wein2022}, we show one can obtain strongly-correlated multi-photon states. Finally, in Section~\ref{sec:mode_selectivity} we also devise a method that enable the emitters interact selectivily with the different topological channels using giant atoms~\cite{kockum18a,Gonzalez-Tudela2019b,FriskKockum2021,Kannan2020,Wang2021a,Giorgi2019QuantumNon-Markovianity}.

\section{Model}
\label{sec:model}

The model that we consider along this manuscript is depicted in Fig.~\ref{fig:1}(a): a quantum emitter interacts locally with one of the edges of a two-dimensional topological insulator. Motivated by recent experiments~\cite{Owens2021}, we particularize for a square photonic lattice with nearest-neighbour hoppings of rate $J$, subject to an effective magnetic flux, $\phi$ --the so-called Harper-Hofstadter (HH) lattice~\cite{hofstadter76a}--, where bands with large Chern number appear for small magnetic fluxes~\cite{Harper2014,Goldman2009}. The bath Hamiltonian then reads (setting $\hbar=1$):
\begin{equation}
H_B=-J\left(\sum_{x,y} a_{x+1,y}^\dagger a_{x,y} +  e^{-2\pi i \phi x}a_{x,y+1}^\dagger a_{x,y}\right) + \text{H.c.} \;,
\end{equation}
where $a^{(\dagger)}_{x,y}$ represent the annihilation (creation) operator at the $(x,y)$ position, and where we take the cavity energy as the energy reference. The emitter is assumed to have a single optical transition between its ground ($g$) and excited state ($e)$ with frequency $\omega_e$, that couples to one of the photonic lattice sites at the edges through the standard light-matter Hamiltonian, $H_I=(g a_{x_e,y_e} \sigma_{eg}+\mathrm{H.c.})$, with $g$ being its coupling strength, $(x_e,y_e)$ the position where it couples, and $\sigma_{\alpha\beta}=\ket{\alpha}\bra{\beta}$ the emitter's operators. The emitter's Hamiltonian then reads, $H_S=\omega_e\sigma_{ee}$, such that the full topological light-matter Hamiltonian reads $H=H_S+H_B+H_I$.\\

\begin{figure}[t!]
\centering
\includegraphics[width=0.9\columnwidth]{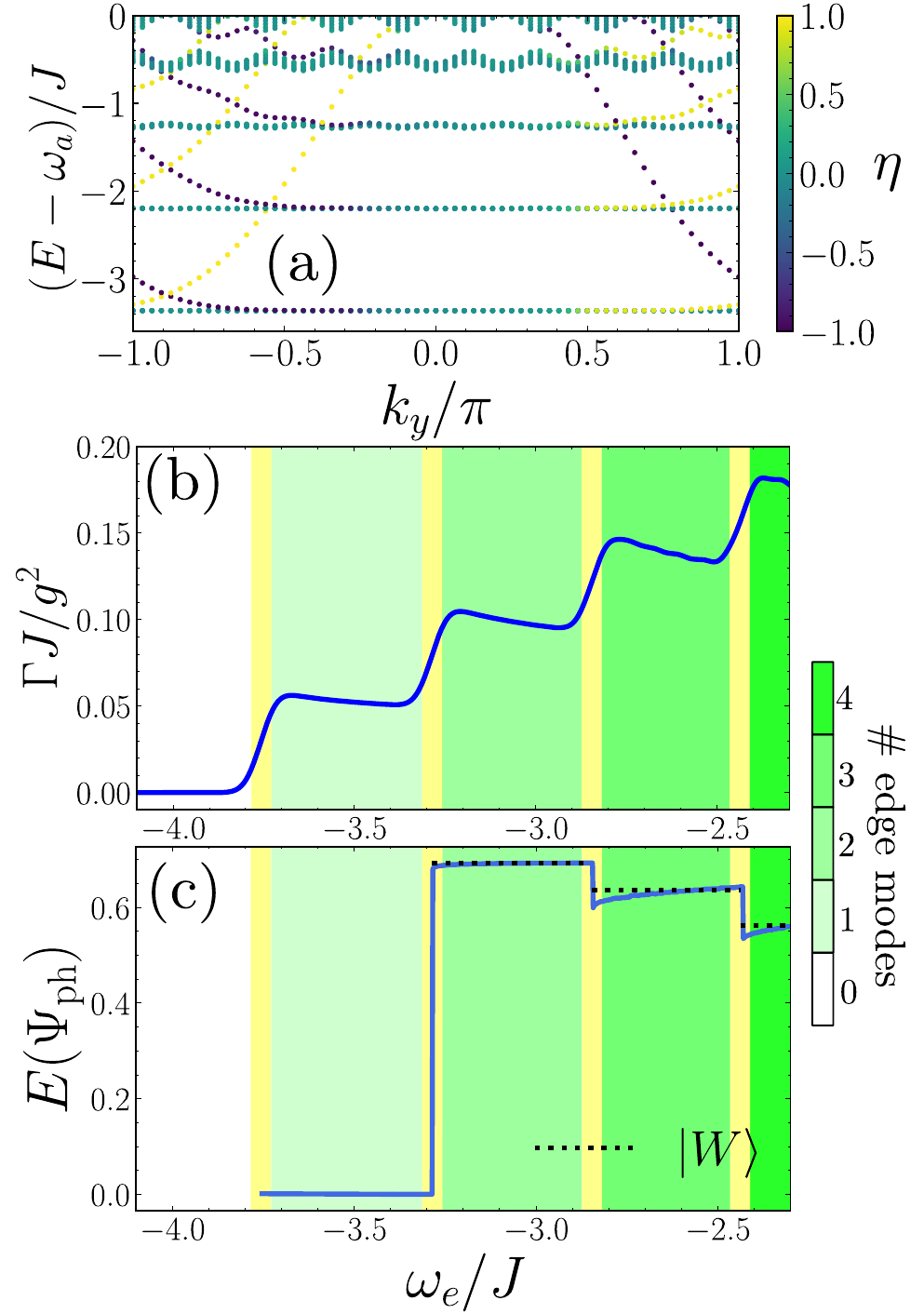}
\caption{(a) HH spectrum with periodic boundary conditions in the Y direction - open for X - for $\phi=1/9$ and a lattice size of $65\times65$ sites. Each dot is coloured according to the localization parameter $\eta$ defined in the main text: $\eta=0$ corresponds to a delocalized state, while $\eta=(-)1$ depicts complete localization at the (left) right boundary. (b) Emitter spontaneous emission rate as a function of its transition frequency $\omega_e$ approximating $\delta(\omega_e-E)$ by a Gaussian function with mean $\omega_e-E_B$ and width $\theta=0.07J$~\cite{SupMatDirectional}. The emitter is coupled to the edge of a HH lattice of size $150\times 150$ with flux $\phi=1/25$. Yellow vertical fringes are centered at Landau levels and have a width equal to $\theta$. (c) Entanglement entropy, $E(\Psi_\mathrm{ph})$, of the emitted single-photon state in the asymptotic limit~\cite{SupMatDirectional}, as a function of the emitter energy, for the same lattice parameters as panel (b).}
\label{fig:2}
\end{figure}

The HH model displays a very rich behaviour depending on the value of $\phi$~\cite{hofstadter76a,Harper2014,Goldman2009}.
here, we take $\phi=1/q$, with $q\in \mathbb{N}$, that is enough to illustrate the behaviour we are interested in. With this parametrization, the spectrum of the system with periodic boundary conditions features $q$ bands, labeled as Landau-levels~\cite{Harper2014}, separated by $l=0,\dots,q-1$ band-gaps. Besides, it can be shown that the first $\left[(q-1)/2\right]$-bands (being $\left[\cdot\right]$ the floor function) have an associated quantized Chern number $C=-1$~\cite{Harper2014,SupMatDirectional}. Thus, with open boundary conditions, it is expected that the $l$-th band-gap features $2(l+1)$ gapless edge states, associated to the $l+1$ Landau-levels below that energy \footnote{As shown in Sup. Material~\cite{SupMatDirectional}, this situation is qualitatively equivalent to the case where the multiple edge states emerge from a single band with $|C|>1$~\cite{Skirlo2014,Skirlo2015}.}. To illustrate this, we consider a cylinder geometry for the bath with periodic (open) boundary conditions in the Y (X) direction. This allows one to write the spatial wavefunction of the bath eigen-states as $\psi_l(x,y)=e^{i k_y y}\Psi_l(x)$, and calculate their eigen-energies $\omega(k_y)$ numerically. In Fig.~\ref{fig:2}(a), we plot an example of the bath spectrum for a bath with $\phi=1/9$ and $65\times65$ sites, showing the emergence of the gapless modes between the bulk flat bands. Besides, we define a localization parameter $\eta = \sum_{x=0}^{L-1} \left(-1+2\frac{x}{L-1}\right)|\Psi(x)|^2$, that ranges $\eta\in [-1,1]$, achieving the extremes $(-) 1$ when the modes are maximally localized at the left (right) edge, encoded in purple (yellow) color, respectively, in Fig.~\ref{fig:2}(a). Like this, one can see how, for energies below $\omega_a$, the edge modes at the left (right) have always negative (positive) group velocity along Y, thus, having a chiral character. Thus, when coupling an emitter to one of the edges, it will only interact with the modes of certain chirality.

An important observation from Fig.~\ref{fig:2}(a) is that the edge modes dispersion deviate significantly from the linear behaviour typically assumed in the literature for such topological channels~\cite{Yao2013,Lemonde2019,Cao2021}. Since this can have important consequences in the quantum optical behaviour, we derive a more accurate effective description of these modes for $\phi \ll 1$, showing the edge-mode dispersion for the left-localized states emerging from the $l$-th Landau Level approximates by~\cite{SupMatDirectional}:
\begin{align}
\omega_{l}(k_y) &\approx \omega_{\mathrm{LL}}(\phi)+ a_l(\phi)(k_y-k_{l}(\phi))^2\,,\label{eq:modeene}
\end{align}
for $(k_y-k_{l}(\phi))\in (-\pi,0)$. Here, $\omega_{\mathrm{LL}}(\phi)/J\approx -4+4\pi\phi\left(l+\frac{1}{2}\right) - \left(\pi\phi\right)^2(l^2+l+1/2)$ is the energy of the $l^\text{th}$-Landau level~\cite{Harper2014}, $a_l(\phi)$ is the effective curvature of the edge modes which we extract from numerical fittings~\cite{SupMatDirectional}, and which converges to $a_l(\phi\rightarrow 0)\sim 0.6$, and $k_{l}(\phi)$ is the momentum resonant to the minimum edge mode energy. Besides, the spatial distribution of these modes along the $X$ direction is $\Psi_{l}(x) = \sqrt{2/\lambda_l(\phi)} e^{-x/\lambda_l (\phi)}$, where $\lambda_l$ grows as $\phi\rightarrow 0$ as expected, since in $\phi=0$ we should recover the delocalized bath eigenstates of the standard square lattice model. 
 
\begin{figure}[t!]
\centering
\includegraphics[width=0.99\columnwidth]{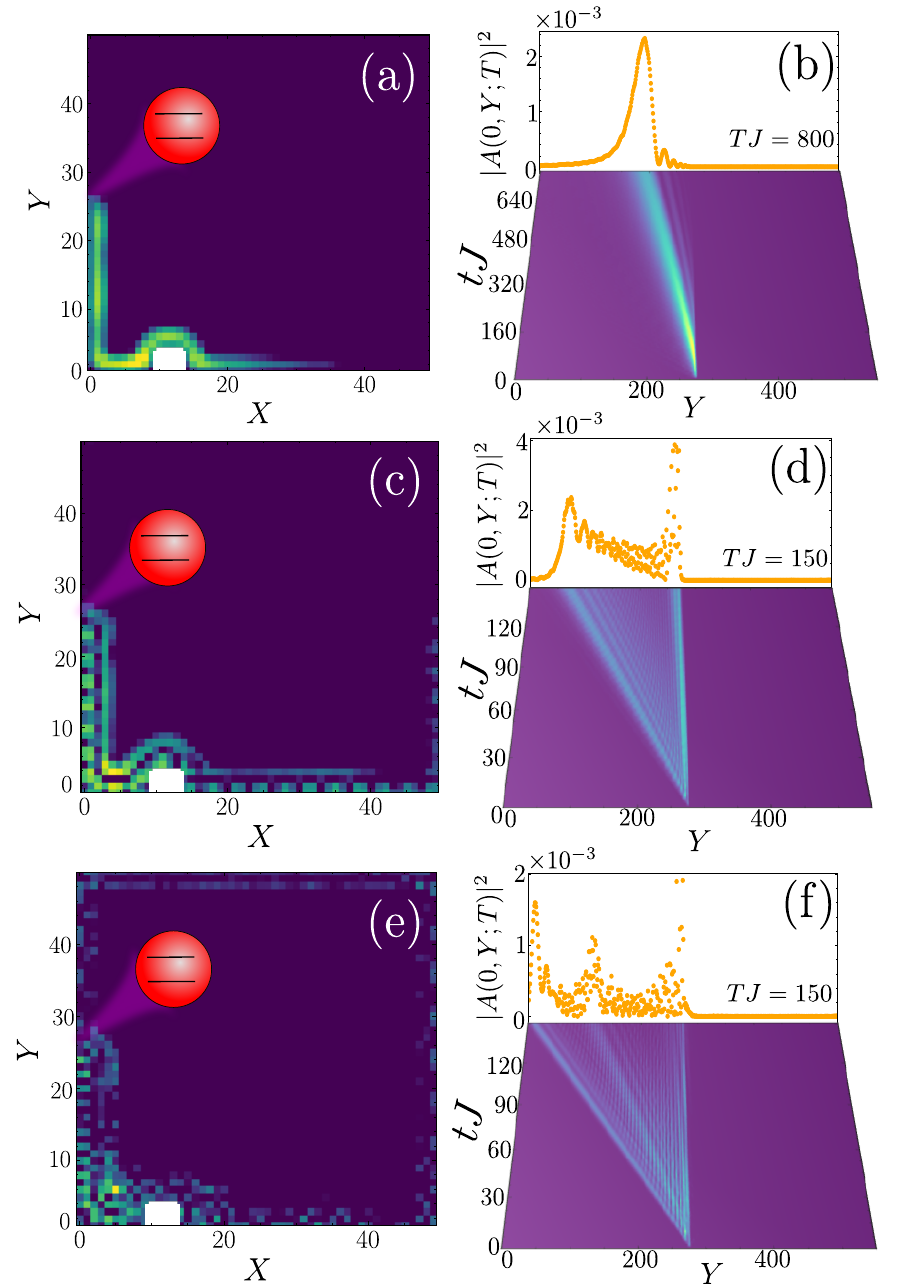}
\caption{Spontaneous emission features of a quantum emitter coupled to the middle site of the left boundary of a HH lattice for $\phi=1/9$, with quantum emitter frequencies of $\omega_e/J=-3.31$ (first row), $\omega_e/J=-2.16$ (second row) and $\omega_e=-1.25$ (last row), resonant to $1$, $2$ and $3$ edge modes respectively. (a, c, e) Snapshots at $TJ=200$ of the bath population in real-space for a lattice size of $50\times 50$ and a coupling constant of $g/J=0.1$, in the presence of a defect, depicted as a white-colored region in the lowest part of the lattice. (b, d, f) Dynamics of the emitted photon. For each $\omega_e$, we plot the evolution of the population of the left boundary sites $|A(0,Y;t)|^2$, in arbitrary units. At the top, we include a snapshot of the pulse shape at the final time instant, showing $1$, $2$ and $3$ peaks respectively, which correspond to the number of resonant edge modes in each case.}
\label{fig:3}
\end{figure}

\section{Spontaneous emission features in multi-mode waveguide QED}
\label{sec:spontaneous_emission}

Let us now see how coupling emitters to the edge of the HH lattice leads to several unique phenomena. First, let us note that since the emitter probes the system at fixed frequency, $\omega_e$, one can control the number of modes that will be relevant for its dynamics just by adjusting its relative detuning with the bath energies. A magnitude that evidences that control is the Markovian decay rate defined by~\cite{CohenTannoudji1998}:
\begin{equation}
\Gamma(\omega_e)=\sum_{E_B}\left|\bra{e}H_I\ket{E_B}\right|^2\cdot\delta(\omega_e-E_B)\;,   
\end{equation}
being $\ket{E_B}$ the bath eigenstates, i.e., $H_B\ket{E_B}=E_B\ket{E_B}$, for the considered configuration. In Fig.~\ref{fig:2}(b), we plot in blue solid line the expected $\Gamma(\omega_e)$ for an emitter coupled to the edge of a HH lattice for $\phi=1/25$ as a function of $\omega_e$. There, we see how the expected decay rate abruptly increases from one band-gap to the other as the emitter's energies is varied. The jumps occur when the emitter's energy $\omega_e$ starts crossing the Landau level energies, indicated in shaded yellow region in the figure, due to the emergence of another edge mode  that couples to the emitter. Note also that the decay rate remains almost constant along the whole band-gap region, except for a deviation that occurs due to the non-linear energy dispersion of the modes. As shown in Ref.~\cite{SupMatDirectional}, this quasi-quantized behaviour is well captured by our effective model, which gives a semi-analytical approximation for the decay rates into the different topological channels $\Gamma_l (\omega_e)$, that reproduces the non-linear dependence with the frequency, i.e.
\begin{equation}
\Gamma_l(\omega_e)\sim(\omega_e-\omega_\text{LL})^{-1/2} \;.  
\end{equation}
As expected, the total decay is then obtained by summing the contributions of the active channels, i.e., $\Gamma=\sum_{l} \Gamma_l$.\\

A more remarkable feature of these topological multi-mode waveguide scenario is what occurs with the spontaneously emitted photons when the emitters are driven. Let us first assume a perfect $\pi$-pulse driving which prepare the system in the state $\ket{\Psi_0}=\ket{e}\otimes\ket{\mathrm{vac}}$, with $\ket{\mathrm{vac}}$ being the bath state with no photons. When the laser is switched off, the whole system evolves according to the total Hamiltonian $\ket{\Psi(t)}=e^{-i H t}\ket{\Psi_0}$, leading eventually to a single-photon wavepacket state
\begin{equation}
\ket{\Psi(t\rightarrow \infty)}=\ket{g}\otimes \sum_{x,y} A(x,y) a_{x,y}^\dagger\ket{\mathrm{vac}}\;,    
\end{equation}
as it occurs in other quantum optical setups~\cite{Law1997DeterministicPulses}. However, in this case such wavepackets have unique features which we illustrate in Fig.~\ref{fig:3} for an emitter coupled to the edge of a HH lattice with $\phi=1/9$. First, irrespective of the band-gap that the emitters are resonant to, the photons are emitted in a chiral and robust fashion due to their topological origin. This is illustrated in Figs.~\ref{fig:3}(a, c, e), where we plot the full photonic bath population, $|A(x,y;t)|^2$, at a time $TJ=200$. There, we observe that the single-photon wavepacket can overcome the defect introduced in one of the edges without altering significantly its propagation. Besides, we also observe how the situations with more than one edge-state, Figs.~\ref{fig:3}(c, e), display a localized, but more complex, wavefunction. To appreciate better the inner structure of these wavepackets, we plot in Figs.~\ref{fig:3}(b, d, f) their temporal dynamics focusing only at the edge population, $|A(0,Y;t)|^2$. Like this, we observe a unique effect of these multi-mode waveguides, that is, after certain time, the emission into the different topological channels becomes spatially separated due to the different group velocities of the modes at the emitter frequency. Intuitively, this separation starts to occur for times such that $|v_{g,l}-v_{g,l'}|T\gtrapprox \Gamma^{-1}_l+\Gamma^{-1}_{l'}$, that is, that the separation between the wavepackets is larger than their intrinsic linewidth ($\Gamma^{-1}_{l'}$).\\

When this separation occurs, one can say that our emitter has generated single-photon entangled states~\cite{VanEnk2005} between orthogonal time-bins $T_l$. Defining $\ket{1}_l$ as the presence of a photon in the $T_l$ time-bin and $0$ in the rest, the photonic state created in the asymptotic limit can be written as
\begin{equation}
|\Psi(t\to\infty)\rangle_\mathrm{ph} \approx \sum_l c_l \ket{1}_l\;.    
\end{equation}
Using that, one can calculate the entanglement entropy~\cite{bennett96a}, $E(\Psi_\mathrm{ph})$, of the asymptotic state in the different band-gaps~\cite{SupMatDirectional} whose result is shown in Fig.~\ref{fig:2}(c) and compared with the one of a perfect $W$-entangled state~\cite{Dur2000} in black dashed line. There, we observe how indeed the entanglement entropy of the emitted state indeed approximates that of maximally-entangled state. Note that in more conventional quantum optical setups~\cite{Gheri1998,Saavedra2000ControlledQubits,Schon2005SequentialStates,Lindner2009,Economou2010,Schwartz2016,pichler17a,Borregaard2020,Wein2022,Tiurev2022,Gimeno-Segovia2019,Kurpiers2019QuantumPhotons,Besse2020,Kannan2020GeneratingElectrodynamics,Wei2021,Wei2022,Ferreira2022DeterministicEmitter} where such time-bin entanglement is generated, it is required to combine superpositions in multi-level emitters and multiple-drivings, while here already with a single $\pi$-pulse, a $W$-type~\cite{Dur2000} entangled structure appears due to the multi-mode nature of the waveguide. Applying several $\pi$-pulses one can obtain more complex multi-photon states. For example, if a second $\pi$-pulse is applied at a time $T_D$ before the excitation from the first $\pi$-pulse decays completely, the emission of the second photon is correlated with the first. As shown in Ref.~\cite{Wein2022}, this ends up generating two-photon Bell-like states in the photon number basis
\begin{equation}
\ket{\Psi_\mathrm{Bell}}\propto (1+a^\dagger_\mathrm{E}a^\dagger_\mathrm{L})\ket{\mathrm{vac}}\;,
\end{equation}
where $a^\dagger_{\mathrm{E}(L)}$ represents the photon operator emitted from the first (second) $\pi$-pulse. Compared to Ref.~\cite{Wein2022}, in the topological multi-mode setups the single-photon wavepackets already have an internal superposition structure, $a^\dagger_\mathrm{E (L)}\propto \sum_l c_{E (L),l} a^\dagger_l\ket{\mathrm{vac}} $, with $a^\dagger_l$ being the effective operator associated to the photon emitted in the $l$-th topological channels. Therefore, its multi-photon structure will be much richer, and depends on the interplay between the pulse delay $T_D$, the global $\Gamma$ and individual $\Gamma_l$ decay times, and the asymptotic time where it is measured. Let us emphasize that the non-gaussian character of these states is a consequence of the strongly interacting character of the emitters, and could never be obtained in classical setups.

\section{Mode Selectivity via non-local couplings}
\label{sec:mode_selectivity}
\begin{figure*}[t!]
\centering
\includegraphics[width=1.9\columnwidth]{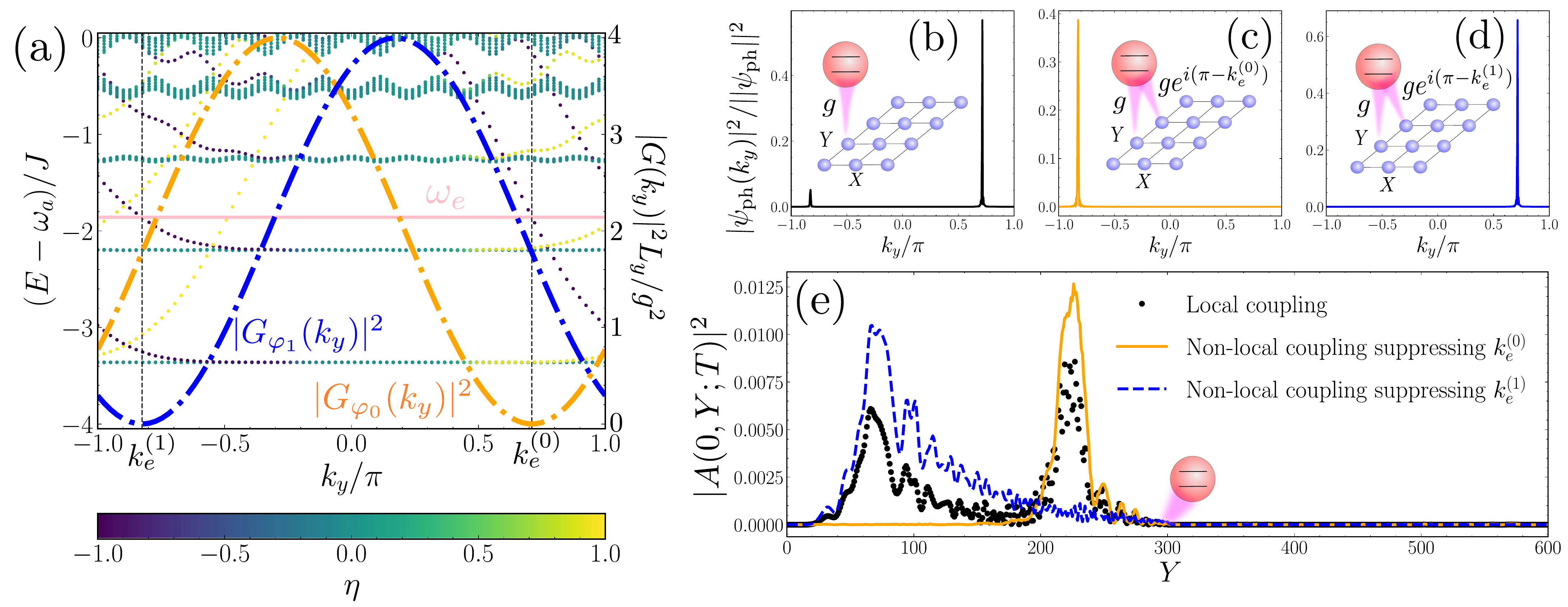}
\caption{(a) Spectrum of a HH lattice of $65\times65$ sites for $\phi=1/12$. Each dot is coloured according to the localization index $\eta$ as in Fig.~\ref{fig:2}(a). The solid pink line depicts the quantum emitter frequency $\omega_e/J=-1.86$, that is resonant to two left-localized edge modes at momenta $k_e^{(0)}$ and $k_e^{(1)}$ respectively. We prove mode-selectivity using non-local couplings that in momentum space are of the form $|G_{\varphi_0}(k_y)|^2$ and $|G_{\varphi_1}(k_y)|^2$ represented as dashdotted lines, which units are indicated in the right vertical axis, and are analytically described in Eq.~\eqref{eq:non_local_coupling}. Note that the orange (blue) line, representing $|G_{\varphi_0}(k_y)|^2$ ($|G_{\varphi_1}(k_y)|^2$) vanishes at $k_e^{(0)}$ ($k_e^{(1)}$). (b, c, d) Normalized photon population in momentum space for the photonic state resulting from spontaneous emission at time $TJ=300$ of a quantum emitter with energy $\omega_e/J=-1.86$ coupled to the left edge of a HH lattice of size $250\times250$ and $\phi=1/12$. The emitter coupling is (b) local (c, d) non-local, designed to cancel emission at $k_e^{(0)}$ and $k_e^{(1)}$ respectively. (e) Snapshot at time $TJ=150$ of the photonic state in the left lattice boundary, coming from spontaneous emission in the three distinct coupling configurations, with same values for $\omega_e$ and $\phi$. The lattice size is $600\times 600$, and the light-matter coupling constant is $g/J=0.2$.} 
\label{fig:4}
\end{figure*}

Finally, let us show how to make the emitters interact selectively with one of the resonant channels~\cite{SupMatDirectional}. The key idea is to couple the emitter with more than a single lattice site, as it can be done with giant atoms~\cite{kockum18a,Gonzalez-Tudela2019b,FriskKockum2021,Kannan2020,Wang2021a}. Let us illustrate it in the simplest case where we want to cancel only one resonant momenta. This requires that the emitter couples to two adjacent cavities with the same strength, and relative phase $e^{i\varphi}$ i.e. $g_{(0,y)}=g$ and $g_{(0,y+1)}=ge^{i\varphi}$. In that case, the $\kk$-dependent light-matter coupling reads
\begin{equation}
|G_\varphi(k_y)|^2\propto \left(1+\cos(k_y+\varphi)\right)\;,
\label{eq:non_local_coupling}
\end{equation}
and thus, vanishes at $k_y=k_e$ if we choose $\varphi=\pi-k_e$. This was the key idea introduced in single mode waveguide QED setups~\cite{ramos16a,Guimond2020ACircuits,Soro2022ChiralAtoms} to obtain chiral emission. Here, the emission is already chiral, but we can still use it to cancel the emission into the resonant momenta of the undesired resonant channels. In Fig.~\ref{fig:4} we show a proof-of-principle realization of that idea for a situation when the emitter is resonant to two edge modes. In Fig.~\ref{fig:4}(a), we plot the energy spectrum for a lattice with $\phi=1/12$ and $65\times 65$ sites. In solid horizontal red line we indicate the energy of the emitter's that is chosen in the second band-gap to be resonant to two channels with resonant momenta $k_e^{(0)}$ and $k_e^{(1)}$, respectively, indicated in vertical dotted black lines. This means that if the emitter couples locally, it will couple to the two $k$-channels, as shown in Fig.~\ref{fig:4}(b), and emit in a two-mode fashion in real space, as shown in black dots in Fig.~\ref{fig:4}(e). On the contrary, if we choose non-local couplings to cancel the coupling to either the momentum $k_e^{(0)}$ or $k_e^{(1)}$, as depicted in orange and blue dotted lines in Fig.~\ref{fig:4}(a), one can see that the emitter selectively emits only in one of the channels, as it is illustrated in Figs.~\ref{fig:4}(c-e) by plotting snapshots of the population in momentum and real space. In particular, in Fig.~\ref{fig:4}(e) where we plot a snapshot at time $TJ=150$ of the spatial profile of the emitted pulse for the different coupling choices, we observe very clearly that the designed non-local couplings suppress the emission onto the selected mode compared to local light-matter coupling situation. In Sup.~Material~\cite{SupMatDirectional}, we also prove that the number of non-local couplings required to cancel $N_k$ resonant momenta scales only linearly with $N_k$. This mode selectivity is an interesting tool in this scenario because when a photon propagates in a chiral channel and interacts with an emitter, it acquires a $\pi$-phase~\cite{lodahl17a}. Thus, if multiple photons are sent in the different channels, such mode selectivity can lead to different phases between the topologically-protected photons. This can be a resource for generating photon gates among topologically-protected photons by adapting existing protocols~\cite{Schrinski2022}.

\section{Conclusions $\&$ outlook}

Summing up, we characterize the topological multi-mode waveguide QED setup that appears when quantum emitters couple to the edges of a Harper-Hofstadter lattice. We find several unique features such as the quasi-quantization of the decay rates and the \emph{spontaneous} generation of entanglement in the different topological channels, as well as a way to make the emitters interact selectively with some of the channels. We foresee that the combination of these setups with multi-level emitters and/or complex time-dependent~\cite{Gheri1998,Saavedra2000ControlledQubits,Schon2005SequentialStates,Lindner2009,Economou2010,Schwartz2016,pichler17a,Borregaard2020,Wein2022,Tiurev2022,Gimeno-Segovia2019,Kurpiers2019QuantumPhotons,Besse2020,Kannan2020GeneratingElectrodynamics,Wei2021,Wei2022,Ferreira2022DeterministicEmitter} or parametric~\cite{mittal16a,mittal18a,Mittal2021,Blanco-Redondo2018,Wang2019b,Blanco-Redondo2020,Doyle2022} drivings can be used to generate more complex states-of-light in these topologically-protected channels either in a transient~\cite{mittal16a,rechtsman16a,mittal18a,Blanco-Redondo2018,Blanco-Redondo2020,Dai2022TopologicallyEmitters,Doyle2022,Tschernig2021TopologicalInsulators} or a steady-state fashion~\cite{Porras2019,Wanjura2020,Ramos2021,Gomez-Leon2021,Gong2018,Kawabata2019,Zirnstein2021,McDonald2018,McDonald2020,McDonald2021}, as well as to induce gates between topologically-protected photons~\cite{Schrinski2022}. Although we find our results for the Harper-Hofstadter model, we expect our findings can be of interest to other system where such multiple edge states appear~\cite{Skirlo2014,Skirlo2015,Trescher2012,Udagawa2014,Bergholtz2015}.\\

\begin{acknowledgements}
  \carlos{The authors acknowledge useful discussions with Geza Giedke and Alberto Mu\~{n}oz de las Heras}. The authors acknowledge support from the Proyecto Sin\'ergico CAM 2020 Y2020/TCS-6545 (NanoQuCo-CM), the CSIC Research Platform on Quantum Technologies PTI-001 and from Spanish project PID2021-127968NB-I00 (MCIU/AEI/FEDER, EU). AGT also acknowledges support from a 2022 Leonardo Grant for Researchers and Cultural Creators, BBVA Foundation. During the writing process of this article, another work studying the coupling of emitters to the edges of two-dimensional photonic insulators appeared~\cite{Zhang2022}, although focused only in the single-edge mode scenario.
\end{acknowledgements}

\section{Appendix}
\setcounter{equation}{0}
\setcounter{figure}{0}
\makeatletter

\appendix
\renewcommand\thefigure{\thesection.\arabic{figure}}    
\setcounter{figure}{0}  
In this Appendix, we provide the details of the calculations supporting the main manuscript. In Section~\ref{secSM:model}, we describe the main characteristic of the spectrum of the topological photonic lattice we consider, explaining how we calculate the spectrum for open and periodic boundary conditions. In Section~\ref{secSM:waveguide}, we focus on the edge modes, and show how their features can be captured by a simple phenomenological model. In Section~\ref{secSM:emitter}, we give more details of the distinctive features of the spontaneous emission of emitters coupled to the edge of these systems, such as their expected decay rates and the spontaneous spatial separation of the photonic emission patterns.

\section{Characterization of the Harper-Hofstadter lattice model~\label{secSM:model}}

The Harper-Hofstadter (HH) model that we consider along this manuscript is a two-dimensional bosonic lattice where time-reversal symmetry is broken by an artificial gauge field introduced through a Peierls phase~\cite{hofstadter76a,Harper2014,Goldman2009}. Denoting by $a^{(\dagger)}_\rr$, the annihilation (creation) operators of the bosonic mode at site $\rr=(x,y)$ of the lattice, the HH Hamiltonian can be written (setting $\hbar=1$ along the manuscript):
\begin{equation}
H_B=-J\left(\sum_{x,y} \left(a_{x+1,y}^\dagger a_{x,y} +  e^{-2\pi i \phi x}a_{x,y+1}^\dagger a_{x,y} \right)+ \text{H.c.}\right) \;,
\end{equation}
where we have assumed that time-reversal symmetry is broken by an artificial uniform magnetic field in the direction perpendicular to the direction of the lattice, which strength is encoded in the value of complex phase $\phi$ acquired in the nearest-neighbour hopping $J$ along the Y direction. Note that we have also dismissed local cavity energy terms $\omega_a\sum_\rr a_\rr^\dagger a_\rr$, assuming all cavity energies to be the same along the lattice $\omega_\rr=\omega_a$, and therefore considering it as the energy reference of the problem setting $\omega_a\equiv 0$. Then, there are three magnitudes that determine the shape of the spectrum and eigenmodes of the system, that are, $J,\phi$, and the system size $L_x\times L_y$.\\

\begin{figure*}[tb]
\centering
\includegraphics[width=1.7\columnwidth]{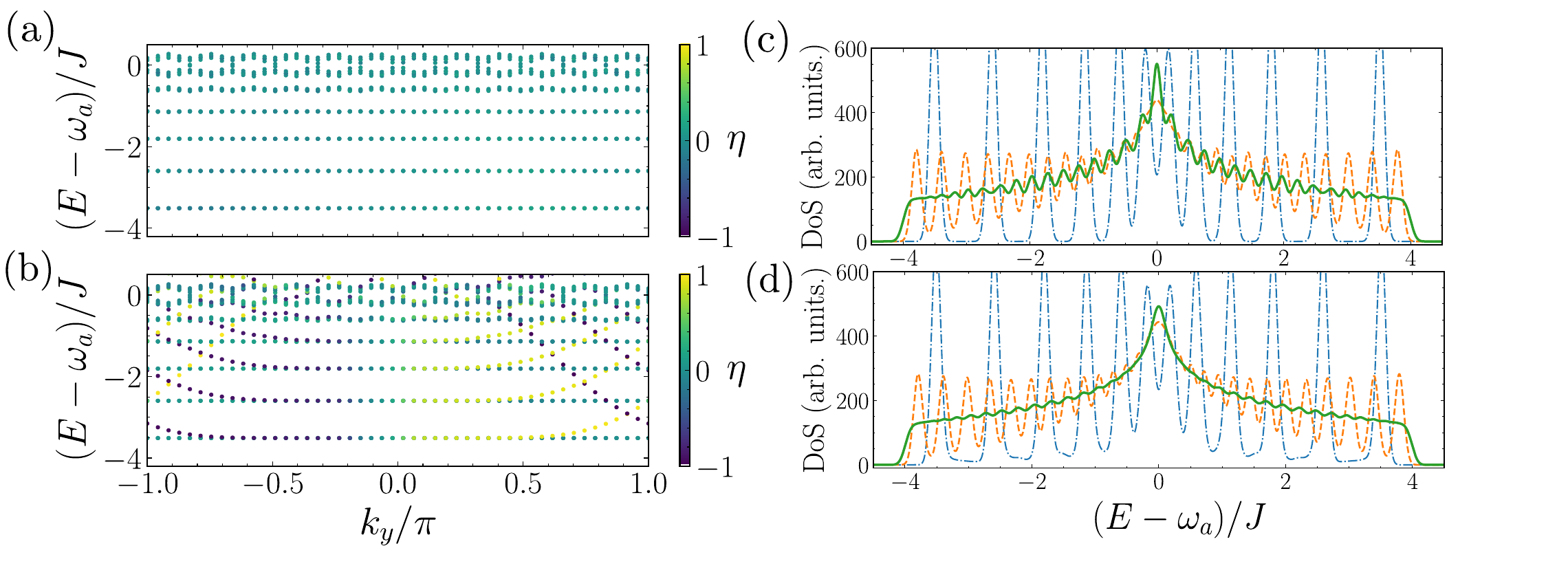}
\caption{ (a, b) Spectrum of a HH lattice of with (a) toroidal and (b) cylinder topology, for a magnetic flux $\phi=1/12$ and a system size of $48\times48$ sites. When periodic boundary conditions are imposed over both spatial directions, panel (a), the lower part of the spectrum consists of a series of flat bands. If a boundary is opened, edge-localized, gapless excitations emerge at band-gap energies, panel (b). The color of each dot is assigned according to the localization parameter $\eta$ defined in Eq.~\eqref{eqSM:loc_index}. (c, d) Density of states (DoS) of a HH lattice as defined in Eq.~\eqref{eqSM:DoS} for the same lattice size, for different values of $\phi$: $1/12$ (dash-dotted blue line), $1/30$ (dashed orange line) and $0$ (solid green line). The topology of the lattice is (c) toroidal and (d) cylindrical, respectively. For $\phi\neq0$, the DoS exhibits divergences at Landau levels. As $\phi\rightarrow 0$, a van Hove divergence is formed at the middle of the spectrum, which is a typical feature of a standard tight-binding square lattice with nearest-neighbor hoppings. The most relevant difference between both instances is that, for cylinder topology, the DoS deviates from zero in the band-gaps due to the presence of topological edge modes. For these fluxes and system size, the ratio between the magnetic lengths $l_B$ and the system size $L=48$ are $l_B/L\approx 0.03$, $l_B/L\approx0.05$ and $l_B/L=\infty$ respectively. Both DoS have been obtained using an auxiliary width (as expressed in Eq.~\eqref{eqSM:normal_function}) of $\theta/J=0.1$.}
\label{figSM:1}
\end{figure*}

Let us start analyzing the spectrum of the system imposing periodic boundary conditions along the X and Y direction, so that momentum $\kk=(k_x,k_y)$ is a good quantum number running over the following values $k_\alpha=-\pi,-\pi+2\pi/L_\alpha,\dots,\pi-2\pi/L_\alpha$. As we said in the main text, we will restrict to rational values of $\phi=p/q$, being $(p,q)$ co-primes. This simplifies the diagonalization of the Hamiltonian, since one can write an effective unit cell that describe the bath lattice of $1\times q$ sites. With that unit cell, and using a plane-wave expansion to account for the periodicity of the lattice, we find that the bath Hamiltonian is given by $q$-bands due to the degeneracy introduced by the super-cell:
\begin{align}
H_B=\sum_{\alpha=1}^q\sum_\kk \omega_\alpha(\kk) c^\dagger_{\alpha,\kk}c_{\alpha,\kk}\,,
\end{align}
where $\alpha$ is the index running over the different bands, and $c^{(\dagger)}_{\alpha,\kk}$ are the operators describing the eigenstates of the bath for a given band $\alpha$ and momentum $\kk$. In Fig.~\ref{figSM:1}(a), we plot an example of that bulk spectrum for $\phi=1/12$ and system size $L_x=L_y=48$. There, we see the emergence of $q$ almost flat bands well-separated in energies. These are the so-called Landau levels that appear in such model~\cite{hofstadter76a,Harper2014,Goldman2009}, and which describe a cyclic motion of lattice excitations whose radius in lattice constant units is given by the \textit{magnetic length}, $l_B = 1/\sqrt{2\pi\phi}$. Thus, in order for these Landau level picture to survive, it is required that such orbit fits in the lattice, i.e., $l_B\ll L_\alpha$. When this condition is not satisfied, the spectrum tends to the square-lattice tight-binding spectrum $\omega(k)=\omega_a-2J(\cos(k_x)+\cos(k_y))$. This transition can be observed in the bath density of states (DoS), defined as
\begin{equation}
\text{DoS}(E) = \sum_{E_B} \delta(E-E_B)\;,
\label{eqSM:DoS}
\end{equation}
where the sum is performed over all the bath eigenenergies $E_B$. Since the Dirac-$\delta$ has only mathematical sense in the continuum limit, for the finite systems we consider we approximate it by a Gaussian distribution of width $\theta$, that is: 
\begin{equation}
\text{DoS}(E) \approx \sum_{E_B} f_\theta(E-E_B)
\end{equation}
with $f_\theta(E-E_B)$:
\begin{equation}
f_\theta(E-E_B)=\frac{1}{\sqrt{2\pi\theta^2}}\exp\left(\frac{(E-E_B)^2}{2\theta^2}\right)\;.
\label{eqSM:normal_function}
\end{equation}

Using that trick, we plot the DoS of the Harper-Hofstadter Hamiltonian in Fig.~\ref{figSM:1}(c), as a function of $\phi$ for a fixed system size, showing the transition from $q$-separated band to a unique band with a Van-Hove singularity at the central energy $\omega_a$, characteristic of the nearest-neighbour two-dimensional tight-binding model.\\

An important characteristic of these energy bands that appear in the bulk spectrum is that they can have a non-zero quantized topological invariant associated to them, that is, the Chern number. In fact, it can be shown that the Chern number of the $l$-th band is given by $C_{l}=t_{l}-t_{l-1}$, where the $t_l$ are integer numbers obtained by solving Azbel-Hofstader Diophantine equation~\cite{Harper2014,Goldman2009}:
\begin{equation}
l+1 = qs_l + pt_l\;,\text{ with }|t_l|<\frac{q}{2}\,,
\label{eqSM:diophantine_eq}
\end{equation}
being $t_{-1}=0$. Notice that the relation $C_{l}=t_{l}-t_{l-1}$ implies that $t_l=\sum_{l'<l}C_{l'}$, i.e. the topological index $t_l$ reveals the sum of the Chern numbers of the lowest $l$ bands. For example, in the case considered in the main text of $\phi=1/q$, it results that the first $\left[(q-1)/2\right]$ bands (letting $\left[\cdot\right]$ be the floor function) will have a Chern number $C_l=-1$. Thus, this choice provides us a way to explore situations with different topological invariants $t_l$ just by probing different energies. Note that other fluxes can lead to different Chern number combinations. For example, as we will see below $\phi=4/9$ leads to a lowest energy band with $C_0=2$, and $\phi=5/14$ leads to $C_0=3$.  

Due to the bulk-boundary correspondence, these quantized topological invariant will have important consequences when we consider open boundary conditions as they will give rise to a number of localized, gapless edge states. To illustrate that, we consider now the system to be placed in a cylinder geometry, such that the system is periodic in the Y direction, but open in the X direction, defining two edges where localized states can appear. To confirm that localized edge states appear, we plot in Fig.~\ref{figSM:1}(b) the spectrum of the system as function of $k_y$, which is still a good quantum number, for the same parameters than the one in Fig.~\ref{figSM:1}(a). There, we observe how on top of the flat bands appearing in the bulk modes, several edge-mode dispersion appears associated to each of the lowest Landau-levels. This makes that the larger the energy of the band-gap, the largest the number of edge states.

To further characterize the properties of the edge modes appearing in such band-gaps, let us note that the eigenmodes in these configuration can be written: $H_B\ket{k_y,\beta}=E_\beta(k_y)\ket{k_y,\beta}$. Projecting their wavefunction into their spatial coordinates $\langle x,y|\Psi_{\beta,k_y}\rangle=e^{ik_y y}\psi_{\beta,k_y}(x)$, one can find that $\psi_{\beta,k_y}(x)$ satisfies the Harper equation:
\begin{align}
\psi_{\beta,k_y}(x+1)+\psi_{\beta,k_y}(x-1)&+2\cos\left(2\pi\phi x - k_y\right)\psi_{\beta,k_y}=\nonumber\\ &=E_\beta(k_y)\psi_{\beta,k_y}(x)\;,
\label{eqSM:edgeequation}
\end{align}
and it features a localized shape. To make it more evident, we define a localization parameter for each eigenstate as follows:
\begin{equation}
\eta = \sum_{x=0}^{L-1} \left(-1+2\frac{x}{L_x-1}\right)|\psi_{\beta,k_y}(x)|^2\;,
\label{eqSM:loc_index}
\end{equation}
which features a maximum $\pm 1$ value when localized in left/right edge, and $0$, when it is delocalized. We codify the value of that parameter in Fig.~\ref{figSM:1}(a) in a color scale where purple/yellow indicates a maximum localization in the left/right edges, whereas blue indicates delocalization. There, we observe another important property of the edge states, that is, that the modes along one edge are perfectly chiral, since they feature a positive/negative group velocity depending on the edge where they are localized. This will have important consequences when an emitter couples to one of the edges, as we will see in Section~\ref{secSM:emitter}. 

\section{Effective edge mode description as a multi-mode waveguide~\label{secSM:waveguide}}

As shown in Fig.~\ref{figSM:1}(b), the first band-gaps of the HH can host a controllable increasing number of edge modes. Since these are effectively one-dimensional modes and chiral, they can be seen as an effective multi-mode one-way waveguide~\cite{Skirlo2014,Skirlo2015}. Generally, such topological have been described within linear approximations~\cite{Yao2013,Lemonde2019,Cao2021}. However, from Fig.~\ref{figSM:1} it is clear that this is not the case in this scenario. In what follows, we will derive a more accurate effective theory that is able to analytically capture the behaviour of these multi-mode waveguides. For concreteness, we will derive such expressions for the left-localized edge states, although a similar description can be found of the right-localized ones.

\subsection{Situation with $\phi=1/q$~\label{subsecSM:1q}}

\begin{figure}[htb!]
\centering
\includegraphics[width=0.89\columnwidth]{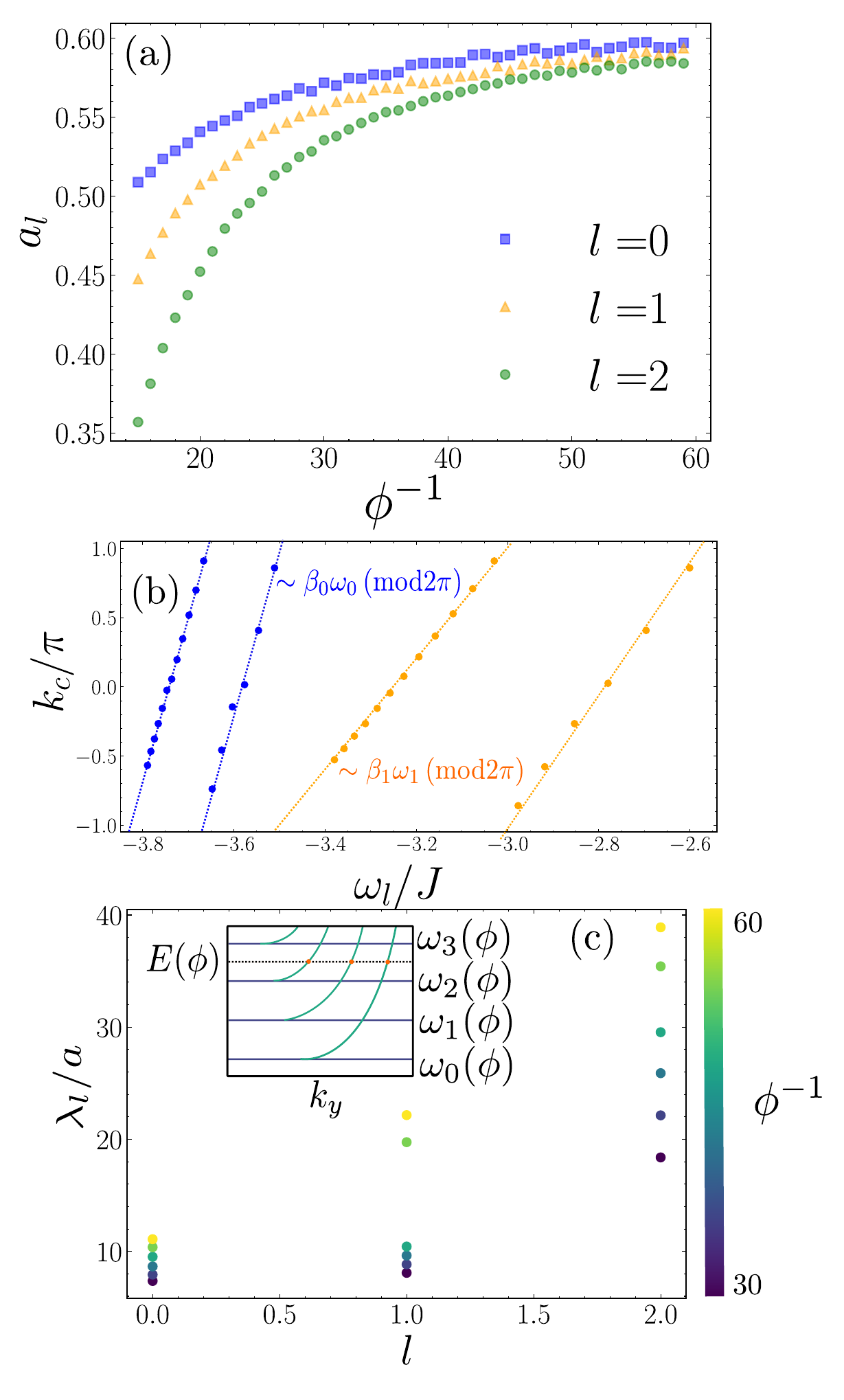}
\caption{(a) Fitted values of the effective model parameter $a_l(\phi)$ in Eq.~\eqref{eqSM:fit}, characterizing the curvature of the edge mode dispersion for the three lowest-energy edge states. In the limit of small fluxes, $\phi\rightarrow0$, all curvatures tend to the same value $\sim0.6$. (b) Fitted values of $k_l$ as a function of the Landau level energy $\omega_l$, for different magnetic flux $\phi=1/q$ ranging from $q=12$ to $q=60$. We observe a linear trend $k_l=\beta_l\omega_l (\text{mod}2\pi)$. Note that the discontinuity between the lines occur because of the definition of $k_l(\pi)$ over the $[-\pi,\pi]$ range. Both figures (a) and (b) are obtained considering a HH lattice of $40\times40$ sites. (c) Localization length $\lambda_l$ of the $3$ lowest edge modes for different values of $\phi$, encoded in the markers' color. All localization lengths are computed at the same energy, namely the middle of the third spectral band-gap, as depicted in the figure inset.}
\label{figSM:2}
\end{figure}

Let us start with the situation $\phi=1/q$ that we consider along the main text and in Fig.~\ref{figSM:1}. After extensive numerical analysis, we find that a good empirical ansatz for the $l$-th eigen-mode dispersion for small magnetic fluxes is given by:
\begin{align}
\omega_{\mathrm{eff},l}(k_y) &= \omega_{l}(\phi)+ a_l(\phi)(k_y-k_{l}(\phi, L))^2\,,\label{eqSM:fit}
\end{align}
where the $\omega_{l}(\phi)$ can be found approximately in the perturbative limit~\cite{Harper2014} as:
\begin{align}
\frac{\omega_{l}(\phi)}{J} = -4 + 2\pi\phi\left(l+\frac{1}{2}\right) - \left(\pi \phi\right)^2\left(l^2+l+\frac{1}{2}\right)+\mathcal{O}\left(\phi^{3}\right)\,,
\end{align}
whereas $a_l(\phi)$ and $k_l(\phi, L)$ are fitting parameters that depend on both the effective flux $\phi$ and $l$-th edge mode considered, although not on system size as long as $l_B\ll L$. Note such quadratic energy dispersions are typical of other (topologically-trivial) waveguides, where the finite size effects introduce energy cut-off for the modes that lead to that behaviour. To further characterize this effective model, we start plotting in
Fig.~\ref{figSM:2}(a) the evolution of the curvature of the edge modes, $a_l(\phi)$, as a function of $\phi$ for the three lowest-energy edge modes in different colors. There, we see how for big fluxes, $\phi$, the curvature of the modes differ significantly, whereas for small fluxes, they converge to a value $a_l(\phi)\approx 0.6$.

Regarding the value of the momentum cut-off $k_l(\phi, L)$, we find that there is a linear dependence with $\omega_l(\phi)$, i.e., $k_l(\phi)=\beta_l\omega_l(\phi)(\mathrm{mod} 2\pi)$. To illustrate that, in Fig.~\ref{figSM:2}(b) we plot one against each other for several fluxes ranging from $\phi\in [\frac{1}{60},\frac{1}{12}]$ in the different markers, together with a dotted line which indicates the result of the fitting. \carlos{To discuss the dependence of $k_l$ on the system size $L$, we may rewrite Eq.~\ref{eqSM:edgeequation} as
\begin{align}
\psi_{\beta,k_y}(x+1)+\psi_{\beta,k_y}(x-1)+V_\phi(x)=E_\beta(k_y)\psi_{\beta,k_y}(x)\;,
\end{align}
where the potential $V_\phi(x)$ is given by:
\begin{equation}
V_\phi(x) = 2\cos\left(2\pi x\phi-k_y\right)\;,\; x=0,1,...,L-1\;
\end{equation}
We observe that varying $L$ modifies the boundary condition of the Harper equation in the right edge. However, if $\phi=1/q$ and $L$ is modified in increased or decreased in $nq$ sites ($n\in\mathbb{N}$), the boundary condition remains, leading to $k_l(\phi, L)-k_l(\phi, L') = 2\pi\phi(L-L')$.}

\begin{figure}[tb]
\centering
\includegraphics[width=0.89\columnwidth]{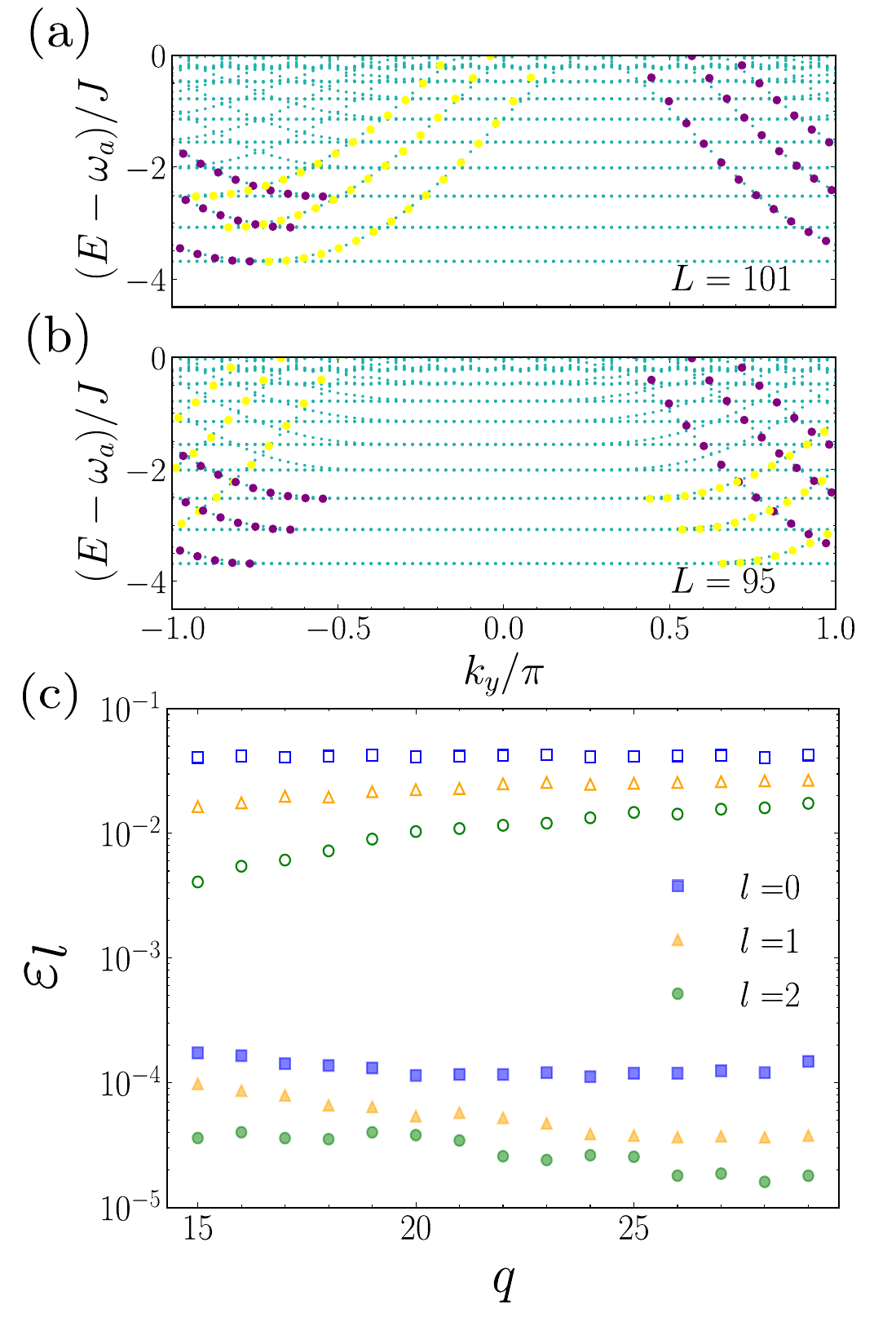}
\caption{(a, b) Effective model vs. exact diagonalization for a magnetic flux $\phi=1/19$ and two different system sizes. In both subfigures, blue dots represent the bath spectrum obtained by exact diagonalization. Purple/yellow dots represent the prediction of the effective model of the energies of edge modes localized at the left/right edge. Notice that varying $L$, the dispersion relation of right-localized states is shifted along $k_y$, which can also be captured by the effective model although we did not explicitly show that. (c) Fitting error, as defined in Eq.~\eqref{eq:error}, for the $3$ lowest edge modes and varying flux. Empty and filled markers correspond to linear and quadratic fittings respectively. We observe that, for all cases, the quadratic fit error is few order of magnitudes lower, showing that this approach is significantly more realistic.}
\label{figSM:3}
\end{figure}

In Fig.~\ref{figSM:3}(a), we show the comparison between the exact diagonalization results and our effective description for a particular value of $\phi$, showing indeed an excellent agreement for the lowest edge states dispersion. In Fig.~\ref{figSM:3}(b), we make a more quantitative assessment on the quality of the model by defining an error parameter:
\begin{equation}
\varepsilon_{l} = \frac{1}{N_k}\sum_{k_y\in \Omega} |\omega_{\mathrm{exact},l}(k_y)-\omega_{\mathrm{eff},l}(k_y)|^2 \;,
\label{eq:error}
\end{equation}
where $\Omega$ is some region in the Brillouin zone along Y where we are interested in performing the approximation to the exactly numerically calculated edge-mode dispersion, $\omega_{\mathrm{exact},l}(k_y)$, and $N_k=|\Omega|$ is the number $k_y$ modes within that region. In Fig.~\ref{figSM:3}(b), we plot $\varepsilon_{l}$ as a function of $\phi$ and compare the accuracy between a linear fit, i.e., $\omega_{\mathrm{eff},l}(k_y)\propto k_y$ (empty markers) and the quadratic fit of Eq.~\eqref{eqSM:fit} (filled markers), showing how indeed the later provides a much more accurate approximation of the modes for all fluxes. 

Apart from the energies, another magnitude of interest of the modes is the localization parameter. In particular, we know from Eq.~\eqref{eqSM:edgeequation} that when $E_{\beta}(k_y)$ lies within a band-gap, the spatial wavefunction along X will be exponentially localized, i.e., $\psi_\beta(x)=\sqrt{2/\lambda_l(\phi)}e^{-x/\lambda_l(\phi)}$, where $\lambda_l(\phi)$ will depend on both the energy level $l$ and the flux $\phi$. In Fig.~\ref{figSM:2}(c) we also plot its dependence, showing that higher-energy edge modes are less localized, and also that increasing the value of $\phi$ yields to higher de-localization.

\subsection{Other situations $\phi\neq 1/q$, $q\in\mathbb{N}$~\label{subsecSM:other}}

\begin{figure*}[tb]
\centering
\includegraphics[width=1.7\columnwidth]{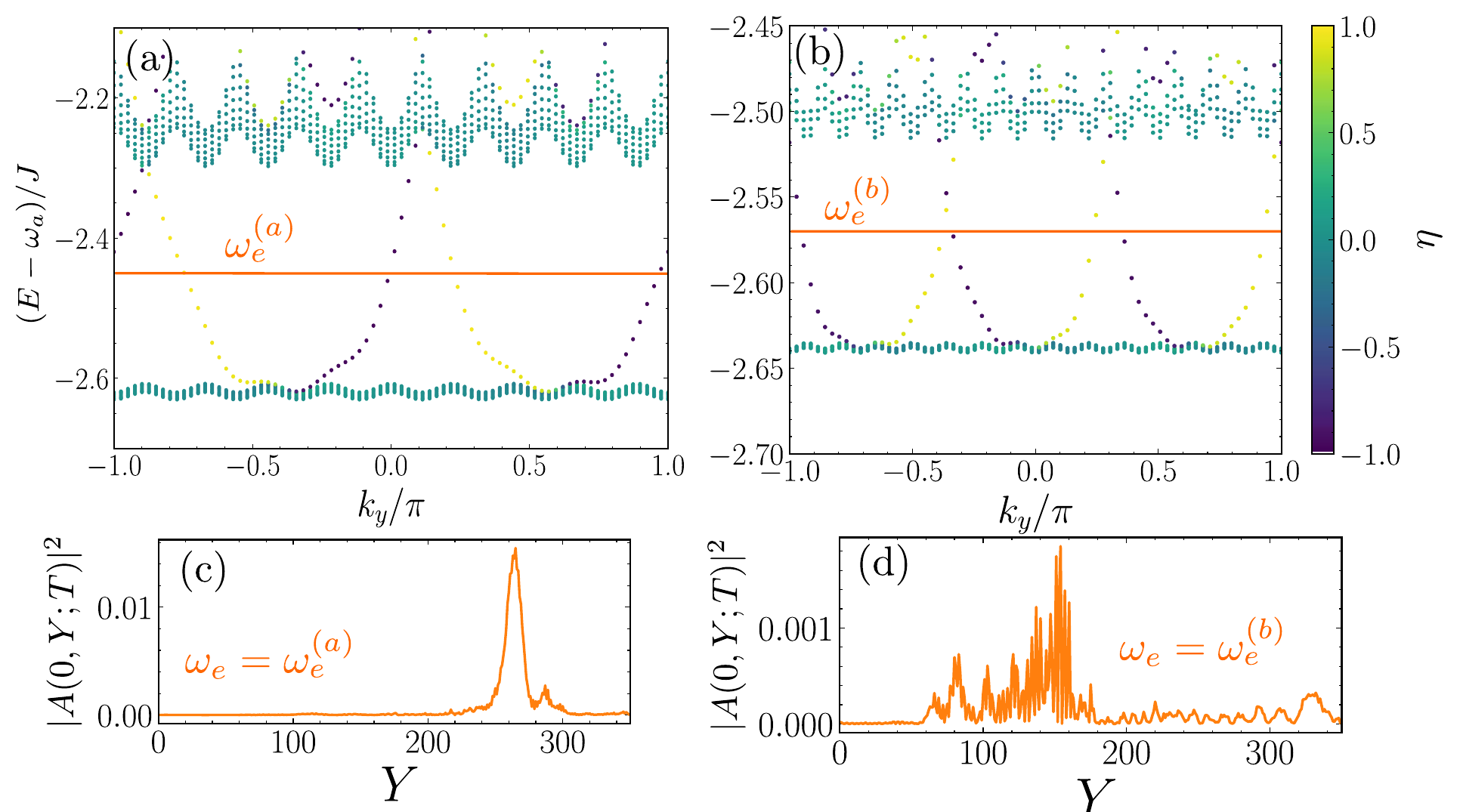}
\caption{(a, b) Lowest energy spectrum of a Harper-Hofstadter lattice of $80\times80$ sites with a magnetic flux of (a) $\phi=4/9$ and (b) $\phi=5/14$. In each case, the lowest band has a Chern number of $2$ and $3$ respectively, which correspond to the number of edge modes per boundary resonant to the lowest spectral band-gap. (c, d) Snapshots of the photonic dynamics at time $TJ=300$ of spontaneous emission along the $Y$ axis of an emitter coupled to the left boundary of the lattice with coupling $g/J=0.2$. The frequency of the emitter in each case is indicated in the (a, b) plots as a solid orange line, and is $\omega_e^{(a)}/J=-2.45$ and $\omega_e^{(b)}/J=-2.57$. We can observe that the pulse shape differs from the single mode scenario. However, it is not possible in this case to resolve the different peaks of the pulse due to the similarity of the group velocities of the different topological channels.}
\label{figSM:4}
\end{figure*}

Along this work, we have restricted to magnetic fluxes of the form $\phi=1/q$, with $q\in\mathbb{N}$, due to the topological features thoroughly discussed along the manuscript, namely the sequence of $\left[q/2\right]$ lowest bands with Chern number $C=-1$. This structure does not prevail if $\phi$ does not fit this form. To illustrate this, we consider the cases of $\phi=4/9$ and $5/14$ in Fig.~\ref{figSM:4}(a, c) and (b, d), which features a Chern-number of $C_0=2,3$, respectively. We start by plotting the spectrum for such values of $\phi$ in panels (a) and (b), where we see that the lowest band-gap feature 4 and 6 edge-state dispersions, respectively, as expected from the value of $C_0$. In general, through numerical inspection we found that in these situations the energy dispersions of the modes tend to be more similar than in the different band-gaps of the $\phi=1/q$ situation. This will result in qualitatively different spontaneous emission patterns, as observed in panels (c) and (d), respectively, where we plot snapshots of the emission in two different situations, illustrating the richness of this setup to obtain qualitatively different photonic wavepackets.

\section{Spontaneous emission of emitters coupled to the edge of the photonic lattice~\label{secSM:emitter}}

In this section, we will consider what happens when a two-level emitter, with Hamiltonian $H_S=\omega_e\sigma_{ee}$, couples to one of the edges of such HH lattice that, for concreteness, we assume to be the left one of Fig.1(a) of the main text. In general, we will consider the most standard local light-matter couplings given by:
\begin{align}
H_I=g\sigma_{eg} a_{\rr_e}+\mathrm{H.c.}\,,\label{eqSM:HI}
\end{align}
where $g$ is the coupling strength of the bath, $\rr_e$ the position of the cavity mode the emitter couples to, and $\sigma_{\alpha\beta}=\ket{\alpha}\bra{\beta}$ the dipole operator of the optical emitter transition that couples the photonic bath. In Section~\ref{subsec:giant}, however, we will consider the non-local couplings that can be engineered with giant atoms~\cite{kockum18a,Gonzalez-Tudela2019b,FriskKockum2021,Kannan2020,Wang2021a}, as a way of selectively coupling some of the topological edge modes.

\subsection{Expected Markovian decay rates or Local density of states~\label{subsec:expected markovian}}

A single emitter can be prepared in its excited state with a classical driving, e.g., using a $\pi$-pulse. If one assumes that the bath has initially no excitations, this state, $\ket{\Psi_0}=\ket{e}\otimes \ket{\mathrm{vac}}_B$, can only evolve into an state of the form:
\begin{align}
\ket{\Psi(t)}=\left(C_e(t)\sigma_{eg}+\sum_\rr A_\rr (t) a^\dagger_\rr\right) \ket{g}\otimes\ket{\mathrm{vac}}_B\,,
\end{align}
because the full light-matter Hamiltonian: $H=H_S+H_B+H_I$, conserves the number of excitations $N_\mathrm{exc}=\sigma_{ee}+\sum_\rr a_\rr^\dagger a_\rr$ since $[H,N_\mathrm{exc}]=0$. Using time-dependent perturbation theory or, equivalently, a Markovian approximation for the system-bath coupling, the emitter is expected to show an exponential decay of its excitation, i.e., $|C_e(t)|^2\approx e^{-\Gamma t}$, with $\Gamma$ being the expected Markovian decay rate given by Fermi's Golden rule~\cite{fermi32a}:
\begin{align}
\Gamma (\omega_e)=&\;2\pi g^2 \sum_\beta |\psi_\beta(0)|^2 \delta\left(\omega_e-E_\beta(k_y)\right) \nonumber\\
&=-2g^2\text{Im}\left[\frac{1}{\omega_e+i0^+-H_B}\right]_{\rr_e,\rr_e}\nonumber\\
&=2g^2\;\text{LDoS}(\rr_e, \omega_e)\,.~\label{eqSM:Gamma}
\end{align}
In the last equality we introduced the local density of states (LDoS) at the emitter position. This quantity is defined similarly to the regular DoS, but weighting the contribution of each bath eigenstate by its support on the position of the emitter:
\begin{equation}
\text{LDoS}(\rr, E) = \sum_{E_B} |\langle \rr|E_B\rangle|^2\cdot \delta(E-E_B)    
\label{eqSM:LDoS}
\end{equation}

\begin{figure}[htb!]
\centering
\includegraphics[width=0.93\columnwidth]{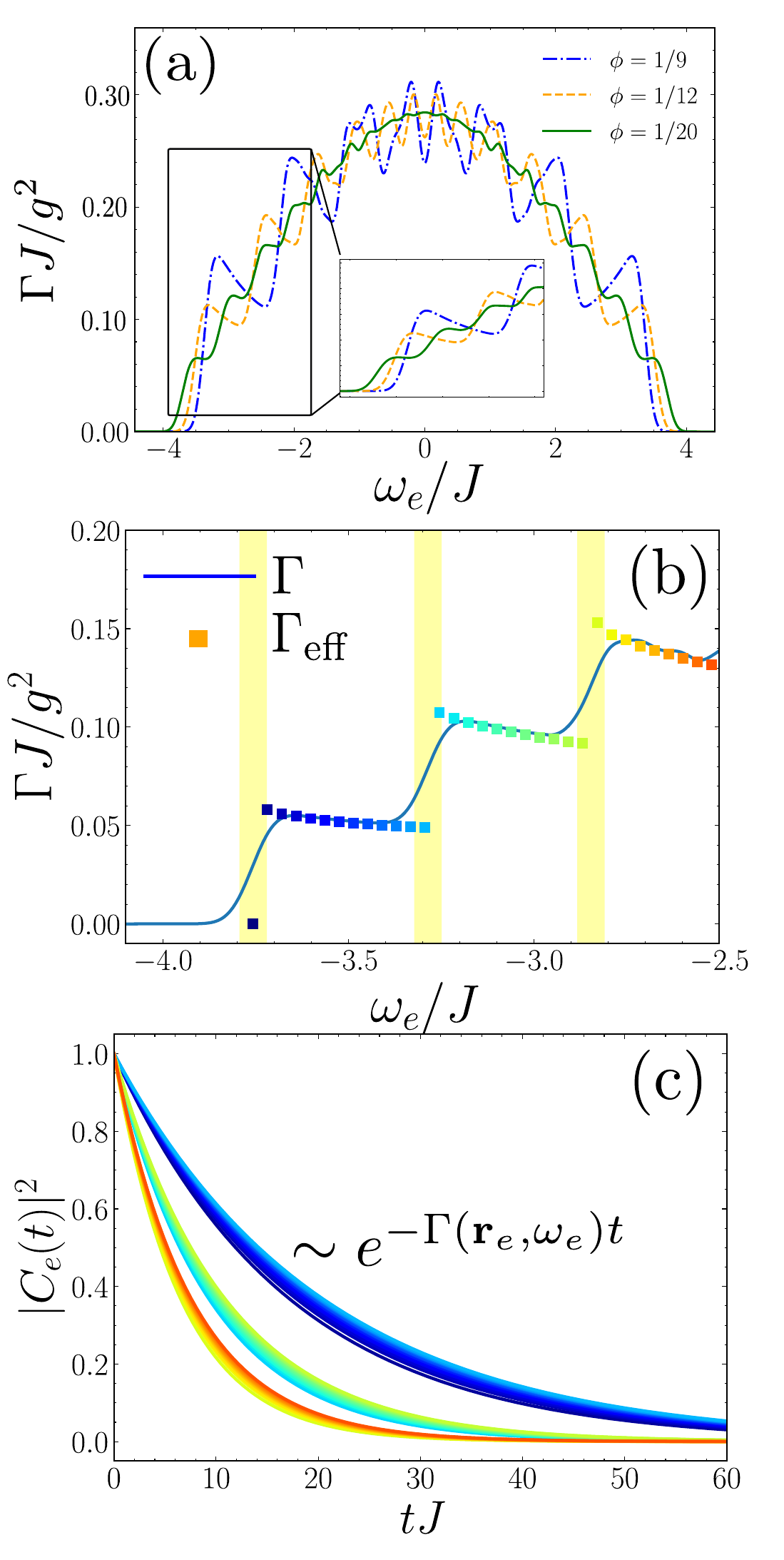}
\caption{(a) Markovian decay rate of a quantum emitter coupled to the edge of a HH lattice of $100\times100$ sites, for different values of the magnetic flux $\phi$, and computed by exact diagonalization (computed using an auxiliary width of $\theta/J=0.15$ as defined in Eq.~\eqref{eqSM:normal_function}). The decay rate curves in the lowest part of the spectrum is zoomed in the figure, showing that lower fluxes support more constant plateaus in the spectral band-gaps. (b) For the same lattice size, the Markovian decay rate for $\phi=1/25$ is depicted as a solid line. Square markers depict the prediction of our effective model for the decay rate in this specific configuration. (c) Spontaneous decay of the emitter population in the same configuration as in (b) for different values of $\omega_e$ and $g/J=0.05$. Each curve is linked to a value of $\omega_e$ and its corresponding color depicted by the square markers in (b).} \label{figSM:5}
\end{figure}

From this definition, the relation of the LDoS with the Markovian decay rate becomes clear by comparing with Eq.(3) of the main text, since
\begin{align}
|\langle e|H_I|E_B\rangle|^2 &= |\langle e|g\sigma_{eg}a_{\rr_e}|E_B\rangle|^2 \nonumber\\
&= g^2|\langle 0|a_{\rr_e}|E_B\rangle|^2 = g^2|\langle \rr_e|E_B\rangle|^2 \;, 
\end{align}
where $|0\rangle=|g\rangle|\text{vac}\rangle$. We numerically compute the LDoS in a similar fashion as we did for the DoS, defining a 'smoothed' Dirac delta function $f_\theta(E-E_B)$, that we take to be a Gaussian distribution as expressed in Eq.~\eqref{eqSM:normal_function}, and computing the LDoS as 
\begin{equation}
\text{LDoS}(\rr_e, E)\approx\sum_{E_B}|\langle \rr_e|E_B\rangle|^2 \cdot f_\theta(E-E_B)\;,   
\label{eqSM:LDoS_approx}
\end{equation}
From Eq.~\eqref{eqSM:Gamma} we read that the shape of the LDoS determines the shape of the expected Markovian decay rate $\Gamma(\omega_e)$. In Fig.~\ref{figSM:5}(a) we plot the Markovian decay rate at the emitter position calculated with exact diagonalization, for several values of $\phi$. We observe that smaller $\phi$'s are associated with narrow band-gaps, situation in which the LDoS 'quasi-quantized' behaviour leads to plateaus in $\Gamma(\omega_e)$. In Fig.~\ref{figSM:5}(b), we represent with a solid line the expected decay rate $\Gamma_e$ as a function of $\omega_e$ for $\phi=1/12$, computed using exact diagonalization, and compare it with the markers that represent the expected decay rate computed as follows:
\begin{equation}
\Gamma_l(\omega_e) \approx \frac{g^2|\psi_l(0)|^2}{|v_{g,l}(\omega_e)|}=\frac{2g^2}{\lambda_l |v_{g,l}(\omega_e)|}\,,  \label{eqSM:gammal}
\end{equation}
where $v_{l,g}(\omega_e)$ is the group velocity of the $l$-th mode at the emitter energy, that is, $v_{g,l}=\partial_{k_y} \omega_{\mathrm{eff},l}(k_y)|_{k_y=k_e}$, with $\omega_{\mathrm{eff},l}(k_e)=\omega_e$. Thus, the total decay rate in a given band-gap will be given by $\Gamma(\omega_e)=\sum_l \Gamma_l(\omega_e)$, where the sum runs over the number of edge modes that are present in that band-gap. We observe that the decay rate displays a ladder-like structure, with jumps located at Landau levels. This quasi-quantization behaviour can be probed by monitoring the decay of the emitter population during spontaneous emission. In Fig.~\ref{figSM:5}(c) we show the quantum emitter dynamics of an initially excited emitter for the range of $\omega_e$ depicted in Fig.~\ref{figSM:5}(b). There, we observe how its timescale remains approximately constant until it crosses the Landau level energy and is able to interact with the new mode of the higher band-gap. We note, however, that the steps of the ladder are not strictly constant due to the non-linear dependence of the mode dispersion. On the contrary, they display the typical $1/\sqrt{\omega-\omega_\mathrm{edge}}$ dependence associated to one-dimensional quadratic band-edge dispersions.\\

\begin{figure}[htb!]
\centering
\includegraphics[width=0.93\columnwidth]{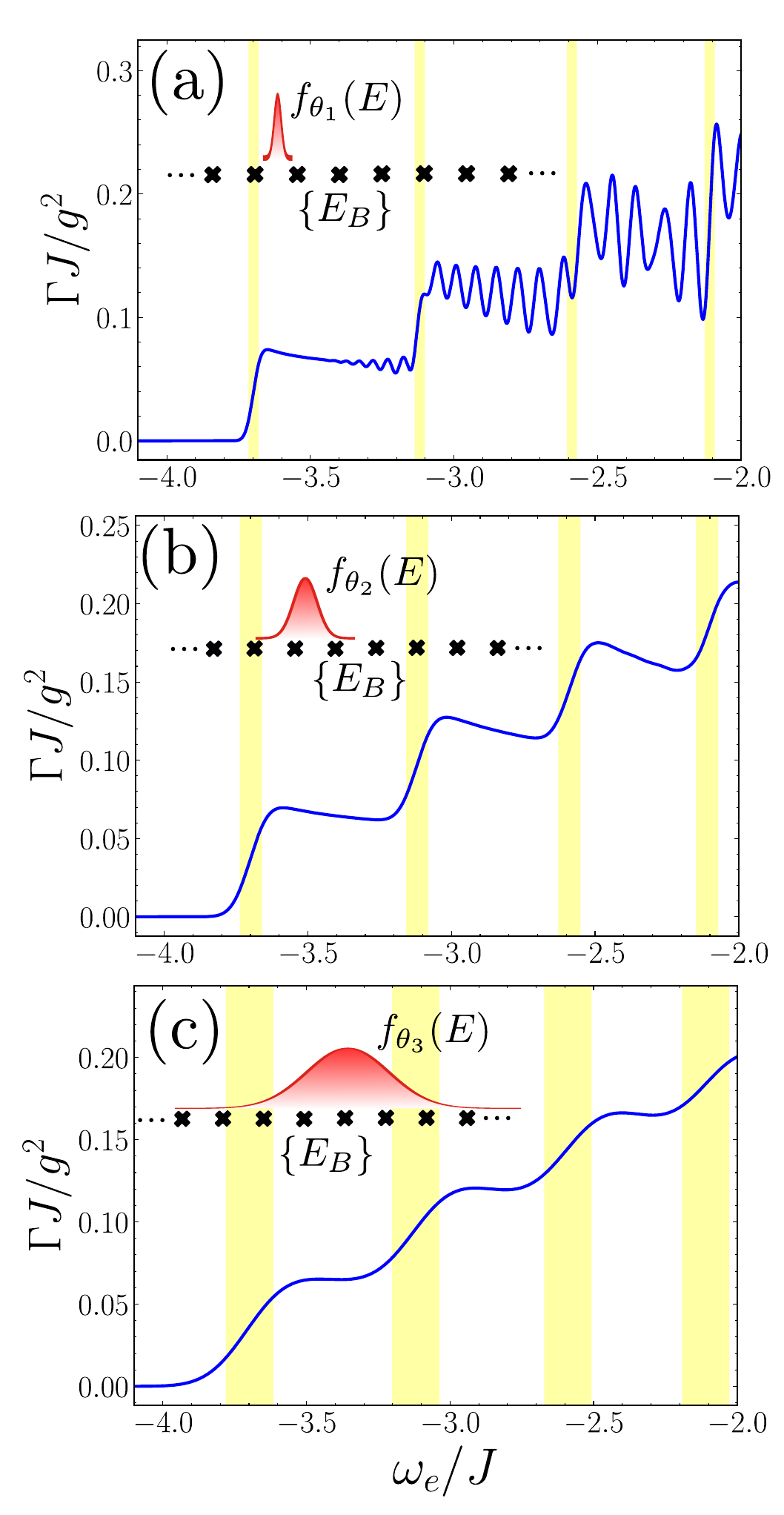}
\caption{Markovian decay rate from exact diagonalization at the boundary of a HH lattice of size $120\times120$ and $\phi=1/15$. The figures are obtained using an auxiliary function width of (a) $\theta_1/J=0.03$, (b) $\theta_2/J=0.07$ and (c) $\theta_3/J=0.16$. The figure insets represent as crossed markers the distribution of bath eigenvalues $\left\lbrace E_B\right\rbrace$. Above these markers, we include a representation of the auxiliary probe Gaussian function $f_{\theta}(E-E_B)$ that is used as described in Eq.~\eqref{eqSM:LDoS_approx} to probe the energy spectrum and compute the LDoS. In each figure, the vertical yellow fringes are centered at Landau levels, and their width is equal to the corresponding value of $\theta$.} \label{figSM:6}
\end{figure}

For the lowest-energy edge modes, we observe the agreement between the our analytical model and the numerical calculations, except near the Landau level divergences. At these energies the numerical computation of the LDoS entails limited resolution, due to the finite width $\theta$ associated to the auxiliary function $f_\theta(E-E_B)$: if we probe the LDoS of an energy $|E-\omega_l|<\theta$ from below, the LDoS will count some states above $\omega_l$, softening the transition. This issue is unavoidable: if we try to have an arbitrarily small value of $\theta$, the probe function will eventually observe the discreteness of the spectrum. In such case, the computation of the LDoS would suffer from numerical instabilities. On the contrary, an excessively value for $\theta$ will make the approximation states in Eq.~\eqref{eqSM:LDoS_approx} increasingly worse. In our case, the information about the LDoS jumps would be lost, softening the shape of the curve. All these numerical issues are graphically represented in Fig.~\ref{figSM:6}.

\subsection{Photonic spontaneous emission patterns~\label{subsec:patterns}}

As we see in the Fig.3 of the main text, the non-uniform group velocity of the modes along a given band-gap favours an spatial separation of the photons propagating into the different channels. This generates naturally single-photon time-bin entangled states~\cite{VanEnk2005}, that when combined with sequential generation methods~\cite{Gheri1998,Saavedra2000ControlledQubits,Schon2005SequentialStates,Lindner2009,Economou2010,Schwartz2016,pichler17a,Borregaard2020,Wein2022,Tiurev2022,Gimeno-Segovia2019,Kurpiers2019QuantumPhotons,Besse2020,Kannan2020GeneratingElectrodynamics,Wei2021,Wei2022,Ferreira2022DeterministicEmitter} can be used to generate complex states of light in these topologically-protected channels. Let us now analyze here the relevant magnitudes that determine such spontaneous separation of the photons focusing on the second band-gap where there are two edge-modes. If we neglect the broadening introduced by the non-linear mode dispersion, which we will see below it is a good approximation in our system, the spatial separation between the different modes is determined by:
\begin{itemize}
    \item The different group velocities, $v_{g,l}$, which makes that after a time $T$, the wavepacket fronts are separated by: $|v_{g,l}-v_{g,l'}|T$. With our choice of units, that length is already normalized to lattice constant units.
    
    \item However, the spatial modes have an intrinsic broadening (in lattice constant units) of order $1/\Gamma_l$ in lattice constant units~\cite{shi15a}.
\end{itemize}

\begin{figure*}[tb]
\centering
\includegraphics[width=1.7\columnwidth]{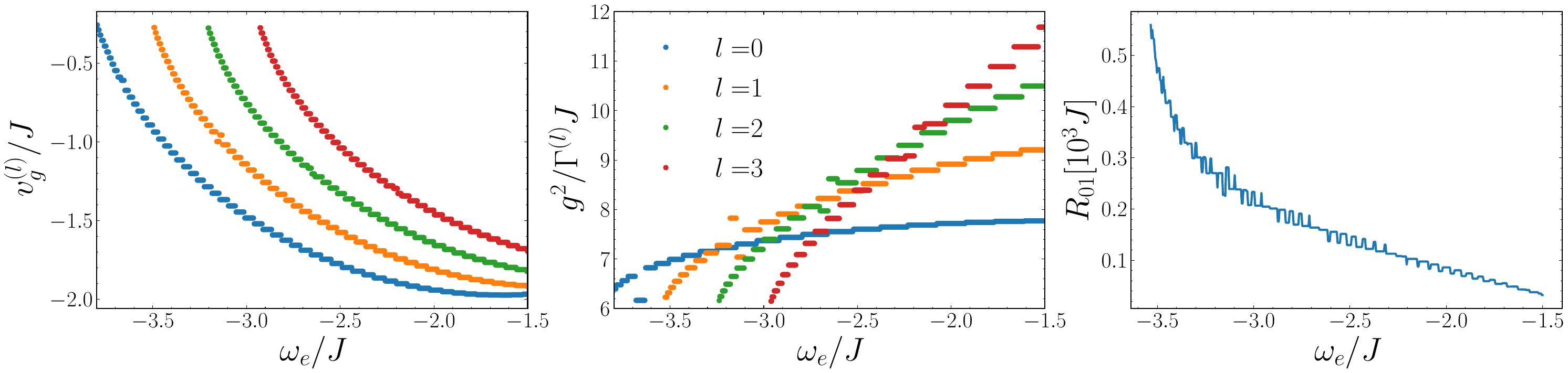}
\caption{(a) Group velocities of the lowest $4$ edge modes of a HH lattice of $50\times50$ sites and $\phi=1/40$. (b) Inverse decay rate $1/\Gamma^{(l)}$ of the quantum emitter onto each topological channel $l$, in units of $J/g^2$, and for the case of $g/J=0.1$. (c) Distinguishability parameter $R_{01}$, as defined in Eq.~\eqref{eqSM:R_coefficient}, for the $2$ lowest edge modes in the $\phi=1/40$ configuration.}
\label{figSM:7}
\end{figure*}

Thus, one can define a parameter $R_{l,l'}$:
\begin{align}
R_{l,l'}=\frac{|v_{g,l}-v_{g,l'}|}{\Gamma_l^{-1}+\Gamma_l'^{-1}}\,.
\label{eqSM:R_coefficient}
\end{align}
which quantifies how favorable is a given configuration to observe the separation of the modes. This quantity has dimensions of energy (i.e. inverse time) and captures the competition between the spatial broadening $\propto \Gamma_l^{-1}+\Gamma_{l'}^{-1}$ and the pulse separation induced by the difference between group velocities $|v_{g,l}-v_{g,l'}|$. When $R_{ll'}T\sim 1$, it is expected that the modes $l$ and $l'$ are fully resolvable at time $T$. In Fig.~\ref{figSM:7} we plot the different group velocities of the modes (panel (a)), broadenings (panel (b)), and  $R_{0,1}$ (panel (c)) as a function of the quantum emitter energy $\omega_e$ for a given bath configuration with $\phi=1/40$.  From Fig.~\ref{figSM:7}(c) we observe that the pulses associated with modes $0$ and $1$ will be fully resolvable at $TJ\sim 10^{3}$. As expected, the best conditions to achieve resolution will occur when the quantum emitter energy $\omega_e$ is slightly above the Landau level energy $\omega_1$: in this case, $|v_{g,0}-v_{g,1}|$ will be maximum, which favours resolution. \\

Apart from the intrinsic broadening of the emitted wavepackets, let us also note that there is an additional source of broadening coming from the curvature of the edge-mode dispersion at the emitter's frequency, that is, $\gamma_\mathrm{dis}$ in $\omega_l(k)\approx \omega_e+v_{g,l}(k-k_e)+\gamma_\mathrm{dis}(k-k_e)^2/2$. In particular, it is well known that a wavepacket with an initial broadening $\sigma_0$ propagating in such a non-linear dispersive channel will have an increasing size growing with:
\begin{equation}
\sigma(t)=\sigma_0\sqrt{1+\frac{\gamma_\mathrm{dis}^2 t^2}{\sigma_0^4}}\;.
\end{equation}

In the case of spontaneous emission of a quantum emitter into the $l^\text{th}$-topological channel of a HH lattice, we can take $\sigma_0$ as the inverse of the decay rate onto such mode, $\Gamma_{l}^{-1}$. The pulse width will then evolve as:
\begin{equation}
\sigma(t)\sim\frac{1}{\Gamma_l}\sqrt{1+\gamma_\mathrm{dis}^2\Gamma_l^4 t^2}\;.
\end{equation}
From this evolution equation, it follows that the role of dispersion will be negligible at a certain time $T$ as long as
\begin{equation}
\gamma_\mathrm{dis}^2\Gamma_l^4T^2\ll1    
\end{equation}
Taking $g/J\sim 0.1$, we find that $\Gamma_l/J\sim 10^{-3}$. On the other hand, we can estimate $\gamma_\mathrm{dis}$ from our effective theory: in particular, we will have that $\gamma_\mathrm{dis}\sim 2a_l(\phi)(k_y-k_l(\phi))$, which will be at most of the order of the unity. Thus, at this value of $g$, dispersion effects will be thus relevant at times $TJ\sim 10^{6}$, which is several orders of magnitude above the timescale where the pulses are separable due to different group velocities, characterized by $R_{ll'}T\sim1$.\\

\subsection{Robustness to disorder~\label{subsecSM:robustness}}

\begin{figure*}[tb]
\centering
\includegraphics[width=1.99\columnwidth]{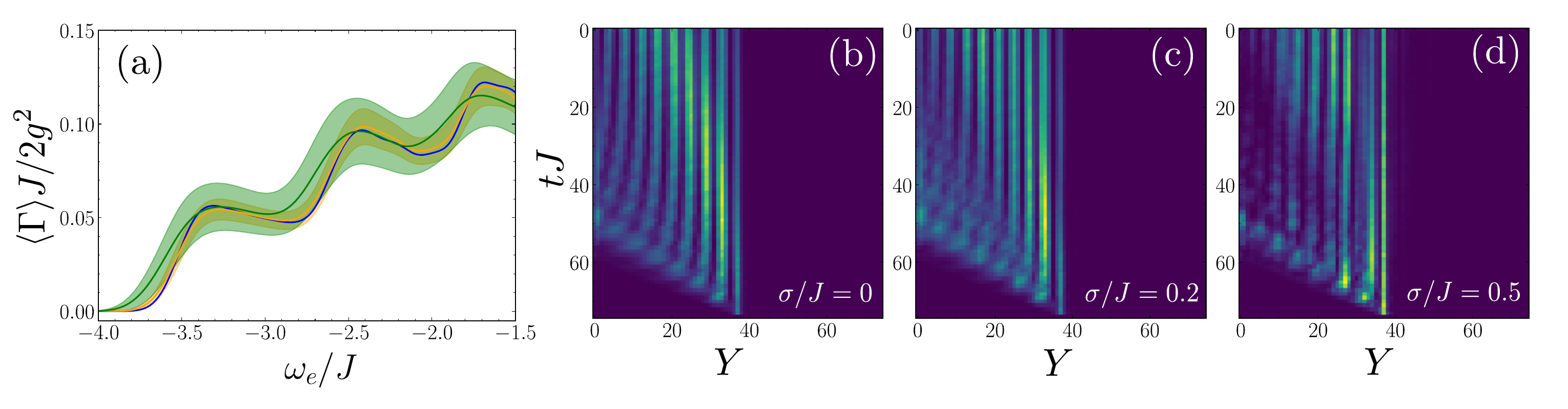}
\caption{(a) Ensemble average of the Markovian decay rate of a quantum emitter coupled at the boundary of a HH lattice of $30\times30$ sites for $\phi=1/12$ for different values of disorder strength: $\sigma/J=0$ (blue), $\sigma/J=0.35$ (orange) and $\sigma/J=0.7$ (green). Each solid line represent the mean LDoS for a collection of $100$ realizations of disorder, which variance is represented as a shaded region around the mean. Every curve is computed by exact diagonalization, using an auxiliary width of $\theta/J=0.15$ as defined in Eq.~\eqref{eqSM:normal_function}. (b)-(c)-(d) Photon transport from the spontaneous emission of such quantum emitter with energy $\omega_e/J=-2.48$ along the left boundary of a HH lattice of $70\times70$ sites and $\phi=1/12$, for several values of disorder strength. We observe that photon transport is robust for small values of $\sigma$. The effects of disorder only start to arise for $\sigma\sim 0.5J$, which is of the order of the band-gap width $E_W\sim0.77J$.} 
\label{figSM:8}
\end{figure*}

In Figs.3(a, c, e) of the main text, we have shown the robustness of single, two, and three-edge mode propagation to an edge defect. This is a consequence of the topological nature of the edge modes. Here, we discuss in greater detail the protection of photon propagation against disorder by introducing random perturbations in the energy of local lattice modes, as follows:
\begin{equation}
H\rightarrow H + \sum_{\rr} \delta\omega_{\rr}a_{\rr}^\dagger a_{\rr}\;,
\end{equation}
where $\delta\omega_{\rr}$ is a random variable uniformly distributed along the interval $(-\sigma, \sigma)$, where $\sigma$ is the strength of the applied disorder. Topological gapless modes spectrally located at a band-gap of width $E_W$ are typically expected to be robust to disorder strengths $\sigma$ up to the order of $E_W$. In Fig.~\ref{figSM:8}, we first analyse the effect of disorder qualitatively in two key features: the LDoS defined in Eq.~\eqref{eqSM:LDoS} and photon emission and propagation. In Fig.~\ref{figSM:8}(a) we plot the Markovian decay rate, computed from exact diagonalization, for an emitter coupled at the boundary of a HH lattice for $\phi=1/12$ for different values of $\sigma$ averaged among different realizations of disorder. We observe that the laddered structure is preserved for values of $\sigma$ comparable to the width of the lowest spectral band-gaps. This is a consequence of the protection of the multi-mode spectrum in the topological band-gaps. Furthermore, in Fig.~\ref{figSM:8}(b-d), we plot photon population emitted for an emitter resonant to the second lowest band-gap of the bath spectrum and for different values of $\sigma$. For weak disorder i.e. small $\sigma$ compared to $E_W$, we observe the same light-cone configuration as in the $\sigma=0$ case. We only start to witness light-cone distortion for values of $\sigma$ comparable to the size of the band-gap that is resonant to the emitter energy. \\

\begin{figure*}[tb]
\centering
\includegraphics[width=1.99\columnwidth]{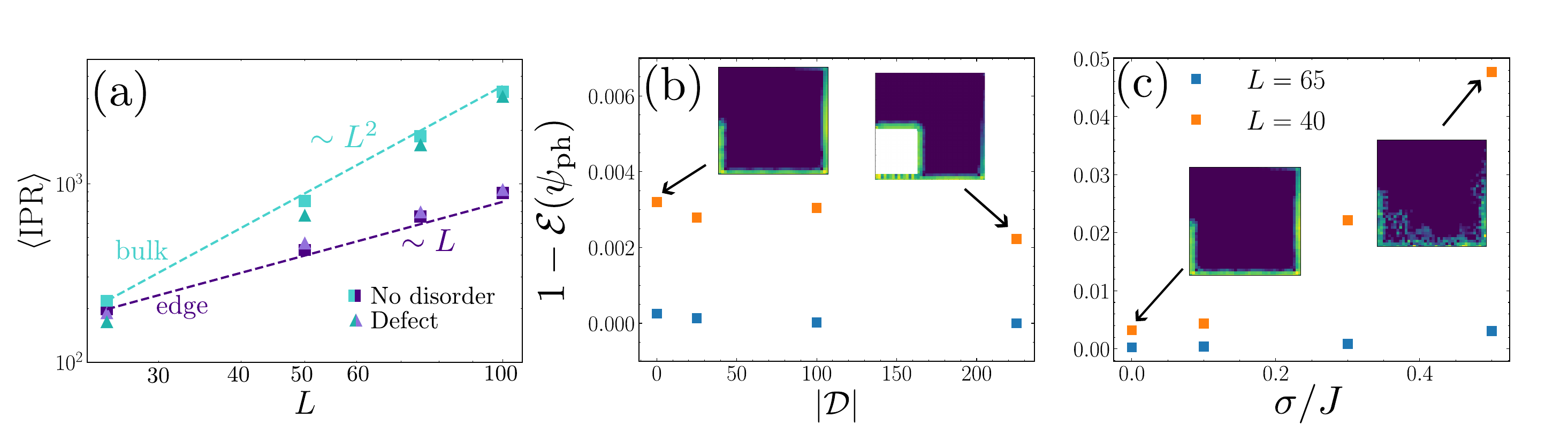}
\caption{(a) Scaling with lattice size $L$ of the Inverse Participation Ratio (IPR) for bulk and edge states in presence vs. absence of defects. The magnetic flux is $\phi=1/9$. The IPR of edge states is computed as the average of all lattice eigenstates with an energy in the lowest spectral band-gap, while bulk states IPR is calculated as an average of eigenstates with an energy equal to the second Landau level. The considered defect has a size of $10\times 10$ sites. (b-c) Effect of disorder induced by defects (b) and local energy perturbations (c) in the edge mode content $\mathcal{E}$ of the spontaneously emitted photon, for two different lattice size. In both cases, we plot $1-\mathcal{E}$ (which can be interpreted as the bulk mode content) for two different lattice size and varying defect sizes and disorder strengths. In both cases, the photonic state is obtained from spontaneous emission of an atom with frequency $\omega_e\approx-2.79J$ and coupled to the lattice with a coupling strength $g=0.05J$, and evaluated at time $T=100/J$. For $L=40$ ($L=65)$, the bulk-edge IPR threshold is set to $500 (800)$. The insets depict snapshots of the spontaneous emission dynamics of an emitter coupled to the center of the left edge, for a lattice size of $L=40$.}
\label{figSM:IPR}
\end{figure*}

\carlos{Now, let us make a more quantitative description of the impact of disorder in the propagation of the edge modes in such multi-mode scenario. For that, we use the method proposed in Ref.~\cite{Tschernig2021TopologicalInsulators}, where they quantify the robustness of photon transport by partitioning the Hilbert space into bulk and edge modes, and observing the edge mode content of the photonic state when defects or local disorder are present. 

To distinguish between edge and bulk modes of the spectrum, we compute the localization properties of each eigenstate of the system using the Inverse Participation Ratio (IPR), defined as follows:
\begin{equation}
\text{IPR}(\Psi) = \frac{1}{\sum_\textbf{r}|\Psi_\textbf{r}|^4}    
\end{equation}
If a wavefunction is spread in a discrete space of $N$ sites, its IPR will be of the order of $N$. Then, we would expect an IPR of the order of $L^2$ for bulk modes (which spread along the whole lattice), and of $L$ for edge states (which are localized along one dimension). In Fig.~\ref{figSM:IPR}(a) we observe these expected scaling relations. The implication of this is that larger lattices leads to larger distinction of the IPR of the bulk and edge modes, thus allowing a clear bi-partition by considering a given treshold $\varepsilon$ in the IPR. After defining the partition of the Hilbert space, we quantify the edge mode content as the associated norm of the emitted wavefunction at a certain time, $\ket{\psi_\mathrm{ph}}$, projected into the edge-mode subspace:
\begin{equation}
\mathcal{E}(\psi_\mathrm{ph}) = \sum_{\mathrm{IPR}(\Psi)<\varepsilon}|\braket{\Psi|\psi_\mathrm{ph}}|^2\;,
\label{eqSM:edge_mode_content}
\end{equation}
where the sum is performed over all lattice eigenstates $\ket{\Psi}$ whose IPR is below a given treshold. In Fig.~\ref{figSM:IPR}(b) we consider a single square defect of increasing size $|\mathcal{D}|$, and study how much the excitation spreads out of the edge modes after impinging with it. We compare the situation with two different system sizes and find that the edge mode content in both cases is practically constant. Furthermore, we see that this robustness measure is higher for larger lattices, where the protection of the edge modes is better. In the inset, we represent a snapshot of the spontaneous emission dynamics where we see that transport is highly protected even for large defects.

On top of that, we also consider a different disorder situation, that is, a random energy disorder over the lattice sites with a normal distribution of width $\sigma$} (see Fig.~\ref{figSM:IPR}c). \carlos{In this case, we find more instructive to project directly to the edge mode subspace of the clean system, since bulk modes undergo Anderson localization, and hence the IPR is no longer valid to differentiate bulk and edge states. In Eq.~\ref{eqSM:edge_mode_content}, this implies doing the sum over the edge eigenstates $\ket{\Psi}$ of the pristine Hamiltonian. In this case, we see that the edge mode content decreases for higher disorder strengths, revealing lack of protection for values of $\sigma$ of the order of $\omega_e-\omega_\mathrm{bulk}$, which is $\approx 0.57J$ in the situation represented in the figure. Again, we also observe that larger sizes favour protection. 
}

\subsection{Photon-loss effects~\label{subsecSM:loss}}

Along all the manuscript, we have neglected the possibility that the emitter or bath modes couples to other additional bath, generating additional decay rates. In that case, the emitter-bath dynamics must be described by a density matrix, $\rho(t)$, formalism. Assuming that the coupling to these baths are Markovian, the dynamics of such density matrix can be described by the following time-local master equation:
\begin{align}
\frac{\partial\rho}{\partial t}=&i[\rho,H]+\sum_\rr \frac{\kappa_\rr}{2}(2a_\rr\rho a_\rr^\dagger-a^\dagger_\rr a_\rr\rho-\rho a^\dagger_\rr a_\rr)\nonumber\\
&+\frac{\Gamma^*}{2}(2\sigma_{ge}\rho \sigma_{eg}-\sigma_{ee}\rho-\rho \sigma_{ee})\,,
\end{align}
where $\kappa_\rr$ and $\Gamma^*$ are the Markovian decay rates induced by these additional baths in the bath and emitter modes, respectively. In the spontaneous emission configuration that we have considered along this manuscript, all the effect of these baths can be captured by replacing the full light-matter Hamiltonian $H$ by a non-Hermitian version which include the effect of the losses:
\begin{align}
H^*=H_S+H_B+H_I-i\sum_\rr \frac{\kappa_\rr}{2}a^\dagger_\rr a_\rr-i\frac{\Gamma^*}{2}\sigma_{ee}\,.
\end{align}

\begin{figure}[tb]
\centering
\includegraphics[width=0.89\columnwidth]{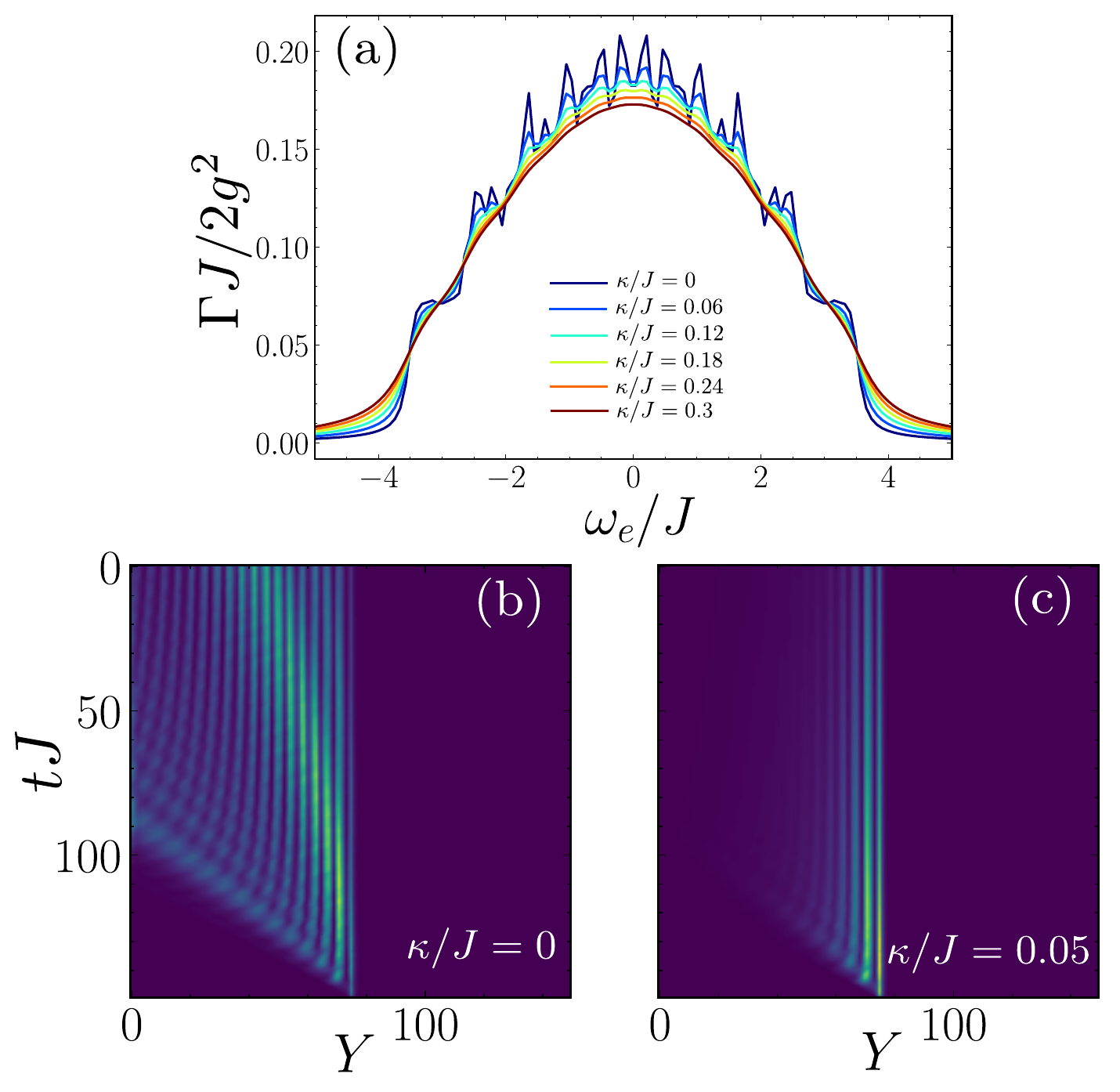}
\caption{(a) Markovian decay rate of a quantum emitter coupled at the edge of a HH lattice of $40\times40$ and $\phi=1/12$, for several values of the local mode loss rate $\kappa$. (b, c) Excitation dynamics along the left boundary of a HH lattice with flux $\phi=1/12$, induced by the spontaneous emission of a quantum emitter with energy $\omega_e/J=-2.48$, resonant to two edge modes, and a local loss rate of (a) $\kappa/J=0$ and (b) $\kappa/J=0.05$ on the lattice sites. We observe two light-cones, associated to the two group velocities of both topological channels.} \label{figSM:9}
\end{figure}

In Fig.~\ref{figSM:9}(a), we plot its effect in the predicted Markovian decay rates for increasing values of $\kappa_\rr=\kappa$ and fixing $\Gamma^*$. There, we observe a ``softening" of the steps of the ladder, although the quasi-quantized behaviour remains unaltered to a great extent. A similar behaviour will occur with $\Gamma^*\neq 0$. The most significant effect of $\kappa$ is to generate a finite propagation length of the photon modes, as observed in Fig.~\ref{figSM:9}(b-c) where we plot an example of the spontaneously emitted photons without and with dissipation, respectively.

\subsection{Single-photon entanglement generation}

In the manuscript, we show that a spontaneously emitted photon from a locally-coupled quantum emitter will be distributed over the multiple boundary modes, and point out that this will lead to a single-particle entangled states between the different channels~\cite{VanEnk2005}. In this section, we give the analytical expression of such photonic state within the Markovian approximation, and quantify its entanglement.\\

Let us first determine the asymptotic photonic state. After a time $t\gg\Gamma^{-1}$, the emitter population will be negligible and the state will be purely photonic i.e. $e^{-iHt}\ket{e}\otimes\ket{\text{vac}}\approx\ket{g}\otimes\ket{\Psi_\mathrm{ph}}$. Assuming a multi-mode waveguide of boundary modes and letting $A_{l,\textbf{k}}$ be the annihilation operator for a photon in the $l^\text{th}$-mode with quasi-momentum $\textbf{k}$, we can rewrite the light-matter Hamiltonian as:
\begin{align}
H=&\omega_e\sigma_{ee}+\sum_l\sum_\textbf{k}\omega_l(\textbf{k})A_{l,\textbf{k}}^\dagger A_{l,\textbf{k}}\nonumber\\
&+\sigma_{eg}\left(\sum_l\sum_{\textbf{k}}g_{l,\textbf{k}}A_{l,\textbf{k}}\right)+\text{H.c.}
\end{align}
with $g_{l,\kk}=\sqrt{\Gamma_l/(2\pi)}$, being $\Gamma_l$ the expected Markovian decay rate in the $\Gamma_l$ mode. Since the Hamiltonian preserves the number of excitations, if the emitter is initially excited, one can write the quantum state of the emitter+ bath system at any time as follows: $\ket{\Psi(t)}=C_e(t)\ket{e}\otimes\ket{\mathrm{vac}}+\sum_\textbf{k}C_\textbf{k}(t)\ket{g}\otimes a_\textbf{k}^\dagger\ket{\mathrm{vac}}$. Applying, the time-dependent Schr\"odinger equation 
to this wavefunction one can obtain the following set of equations for the time-dependent coefficients:
\begin{align}
\frac{\partial C_e(t)}{\partial t}&=\sum_{l}\sum_\textbf{k}ig_{l,\textbf{k}}e^{i(\omega_e-\omega_l(\textbf{k})t)}C_\textbf{k}(t)\nonumber\\
\frac{\partial C_\textbf{k}(t)}{\partial t} &= \sum_{l}ig^\star_{l,\textbf{k}}e^{i(\omega_l(\textbf{k})-\omega_e)t}C_e(t)
\end{align}
Formally integrating the second equation, inserting in the first equation and applying Markov approximation, we find that the emitter follows exponential decay $C_e(t)\approx e^{-\Gamma t/2}$, where $\Gamma=\sum_l\Gamma_l$. Then:
\begin{align}
C_\textbf{k}(t) &\approx \sum_l\int_0^t d\tau\; ig_{l,\textbf{k}}e^{i(\omega_l(\textbf{k})-\omega_e)\tau}e^{-\Gamma \tau/2}\nonumber\\
&\approx\sum_l \sqrt{\frac{\Gamma_l}{2\pi}}\frac{e^{i(\omega_l(\textbf{k})-\omega_e)t}}{\omega_l(\textbf{k})-\omega_e+i\Gamma/2}    
\end{align}
where we assumed $t\gg \Gamma^{-1}$. We conclude that the asymptotic photonic state $\ket{\Psi_\mathrm{ph}}$ can be written as
\begin{equation}
\ket{\Psi_\mathrm{ph}} = \sum_{l, \textbf{k}} \sqrt{\frac{\Gamma_l}{2\pi}}\frac{e^{-i\omega_l(\textbf{k})t}}{\omega_l(\textbf{k})-\omega_e+i\Gamma/2}\ket{n_l\textbf{k}}
\end{equation}

For a single mode, we recover the well-known expression for the asymptotic photonic state of the emitted light from an emitter coupled to a single-mode waveguide~\cite{shi15a}. From this expression, we observe that the contribution of the $l^\text{th}$-mode to the asymptotic photonic state population is weighted by $\sqrt{\Gamma_l}$. Defining a multi-mode basis $\left\lbrace \ket{1_l}\right\rbrace$, where $\ket{1_l}$ contains one photon in the $l^\text{th}$-mode, and defining $\mathcal{C}_l=\sqrt{\Gamma_l/\Gamma}$ we can write the photonic state whose entanglement we want to characterize as:
\begin{equation}
\ket{\Psi_\mathrm{ph}}=\sum_l \mathcal{C}_l\ket{1_l}\,.
\label{eq:statesimp}
\end{equation}

\begin{figure}[tb]
\centering
\includegraphics[width=0.99\columnwidth]{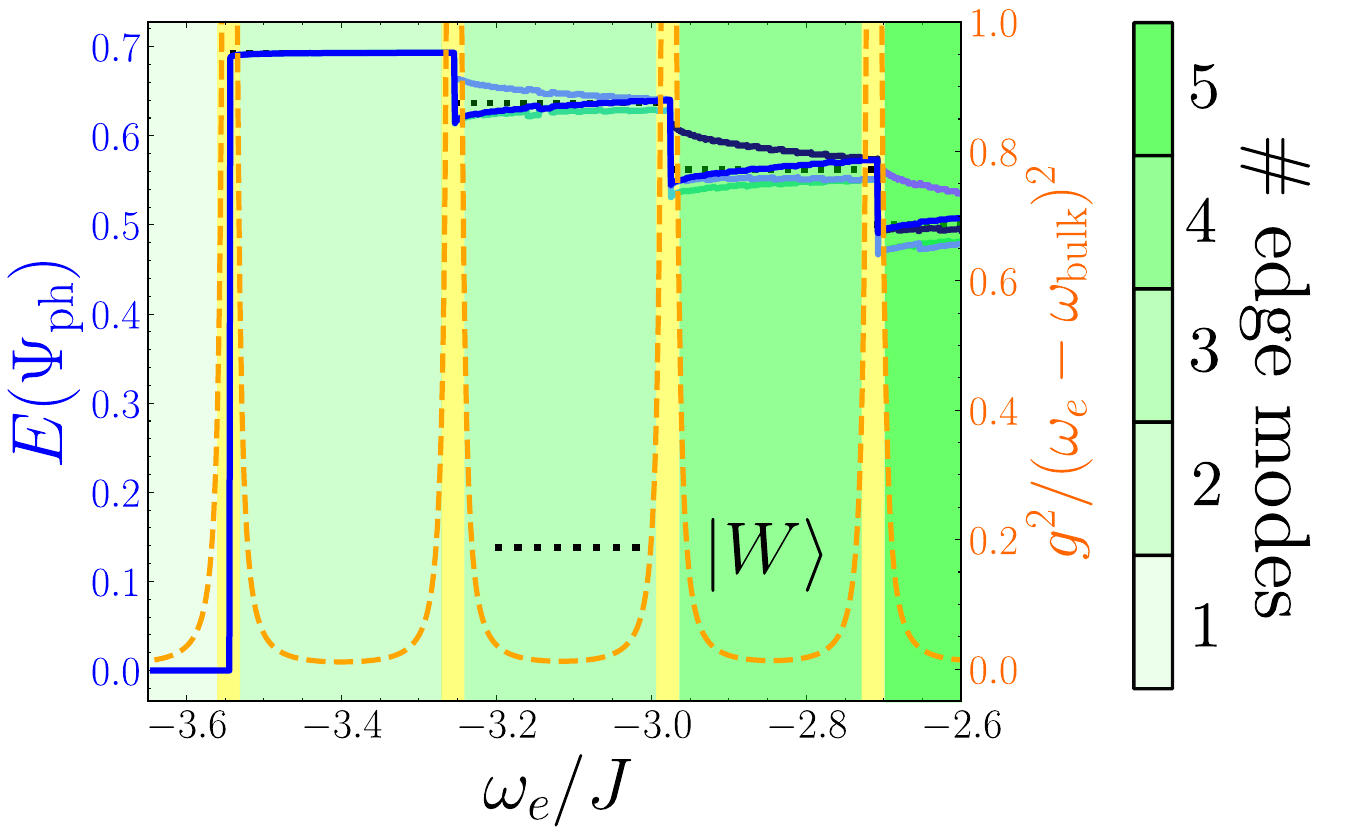}
\caption{\carlos{Quantifying single-particle entanglement of spontaneously emitted photon states. For a Harper-Hofstadter lattice of $\phi=1/25$, we represent the entanglement entropy for all possible bi-partitions (blue solid lines) and the approximate population of the bulk modes obtained using a perturbative approximation (orange dashed line) for different values of the emitter transition frequency. Yellow regions depict the location of the Landau levels. Horizontal black dotted lines represent the entanglement entropy benchmark for $W$-states with a number of qubits equal to the number of resonant edge modes in each band-gap.}}
\label{figSM:entanglement}
\end{figure}

Here, we are consciously neglecting the contribution of the different phases between the modes, since in principle they can be corrected by local operations, and thus should not contribute to the entanglement measures. One way of characterizing the entanglement of this class of states is by calculating the entanglement entropy~\cite{bennett96a} between the different topological channels.  The entanglement entropy of a given state $\Psi$ for a bipartition of the Hilbert space $\mathbb{H}_A\otimes\mathbb{H}_B$ reads~\cite{bennett96a}:
\begin{equation}
E\left(\ket{\Psi}\bra{\Psi}\right) \equiv S\left(\text{Tr}_A \ket{\Psi}\bra{\Psi}\right)  =S\left(\text{Tr}_B\ket{\Psi}\bra{\Psi}\right)                \;,
\label{eqSM:entanglement_entropy}
\end{equation}
where $S(\rho)=-\text{Tr}(\rho\;\log\rho)$ and $\text{Tr}_{A/B}$ denotes the partial trace along the $A/B$ subspace. In our case, we define as $P_l$ to the bi-partition separating the $l^\text{th}$-mode from the rest. Assuming a bi-partition $P_l$, we will take our Hilbert space as $\mathbb{H}_l\oplus \left(\oplus_{l'\neq l}\mathbb{H}_{l'}\right)$,  and denote by $\ket{10}$ ($\ket{01}$) the state with one excitation in the $l^\text{th}$-mode and zero in the rest (and viceversa). From Eq.~\eqref{eq:statesimp}, and using that notation, we then know we can always write our state as:
\begin{equation}
\ket{\psi_\text{ph}} = \mathcal{C}_l\ket{10}+\sqrt{1-|\mathcal{C}_l|^2}\ket{01}\;.
\end{equation}
Denoting by $p=|\mathcal{C}_l|^2$ the probability of measuring the photon in the $l^\text{th}$-mode, the entanglement entropy of this state can be readily computed as 
\begin{equation}
E(\ket{\psi_\mathrm{ph}}\bra{\psi_\mathrm{ph}})=-p\log p-(1-p)\log(1-p)    \;\label{eq:entangbi}
\end{equation}
In Fig.~\ref{figSM:entanglement}, we present the entanglement entropy for all possible bi-partitions in each spectral band-gap, in comparison with the entanglement entropy of a $W$-state with a number of qubits equal to the number of resonant edge modes in each band-gap (dashed lines). The latter can be calculated by imposing $p=1/N_\mathrm{modes}$ in Eq.~\eqref{eq:entangbi}. In the figure, we observe that the entanglement of our generated state is very similar to the one obtained for a maximally-entangled W state in all the band-gaps. Let us note that this calculation is assuming that the emitter only decays through the edge modes. This will be a good approximation as long as the population of the bulk modes remains small. The latter can be perturbatively approximated by $g^2/(\omega_e-\omega_\mathrm{bulk})^2$, which is what we plot in dashed orange curves in Fig.~\ref{figSM:entanglement}. Thus, the calculation of the entanglement entropy will be valid outside of the yellow regions where the population into the Landau levels become negligible. To make a more quantitative estimation of the impact of the bulk-modes into the entanglement generated, one can use more sophisticated entanglement measures such as the negativity~\cite{vidal2002} that can be effectively computed for mixed states. However, we believe that for the sake of illustration the entanglement entropy between the different channels represent a more intuitive witness.\\

\subsection{Mode-selectivity using non-local couplings~\label{subsec:giant}}

Local light-matter Hamiltonians, as the ones considered in Eq.~\eqref{eqSM:HI}, lead to completely delocalized couplings in momentum space. For example, assuming the cylinder geometry that we considered in the previous section, one can rewrite the light-matter Hamiltonian $H_I$ as follows:
\begin{align}
H_I&=g\sigma_{eg} a_{\rr_e}+\mathrm{H.c.}\nonumber\\
&=\sum_{k_y}\left(\frac{g}{\sqrt{L_y}}\sigma_{eg}a_{(x_e,k_y)}e^{-i k_y y_e}+\mathrm{H.c.}\right)\,,\label{eqSM:HI2}
\end{align}
where:
\begin{align}
   a_{(x_e,k_y)}=\frac{1}{\sqrt{L_y}}\sum_{y} a_{(x_e,y)}e^{-i k_y y}\,.
\end{align}

This makes that when coupling an emitter with an optical transition of a given energy, $\omega_e$, it couples to all the edge modes of a given energy. Thus, a quantum emitter locally coupled to the edge of a HH lattice will couple to all topological channels. In this section, we address the problem of selecting the channels onto the quantum emitter can decay by using non-local couplings such that the emitted photon couples selectively to some of the topological channels. The key point is that if one let the emitter to couple several edge sites with different strength, $g_{j}$, the light-matter coupling acquires a $\kk$-dependence depending on these couplings:
\begin{align}
H_I=&\sum_y g_y \sigma_{eg} a_{(x,y)}+\mathrm{H.c.}\nonumber\\
=&\sum_{k_y}\left(G(k_y)\sigma_{eg}a_{(x_e,k_y)}+\mathrm{H.c.}\right)\,,\label{eqSM:HI22}
\end{align}
with:
\begin{align}
G(k_y)=\frac{1}{\sqrt{L_y}}\sum_y g_y e^{-i k_y y}\,.\label{eq:Gk}
\end{align}

Such $\kk$-dependent coupling can be used to cancel the coupling to certain momentum and thus prevent the emission in these modes. For example, this has been used in one~\cite{ramos16a} and two-dimensional~\cite{Gonzalez-Tudela2019b} single-mode waveguides to generate chiral, one-directional emission in baths with isotropic bath dispersions. In our topological multi-mode waveguides, for fixed $\omega_e$, the resonant modes feature a different resonant $k_{e}^{(l)}$.  Thus, choosing the $g_y$ such that $G(k_{e}^{(l)})=0$ to some of the modes, one can selectively couple to the other ones.\\

Let us now illustrate a potential method to achieve that selectivity inspired by the results in Refs.~\cite{ramos16a,Gonzalez-Tudela2019b} that only requires coupling to $N_c+1$ cavities if one wants to cancel the coupling to $N_c$ resonant momenta $k_{e}^{(l)}$. Let us start by the case where we want cancel only the coupling to a single resonant momenta $k_{e}^{(l)}$. For that, let us assume that the emitter couples to neighbouring sites with the same amplitude, but a relative complex phase $\varphi_l$, that is: $g_{y}=g$, $g_{y+1}=g e^{i\varphi_l}$. Using these values, it can be shown from Eq.~\eqref{eq:Gk} that:
\begin{align}
    |G_{\varphi_l}(k_y)|^2=\frac{2g^2}{L_y}(1+\cos(k_y+\varphi_l))\,.\label{eqSM:Gphik}
\end{align}

Thus, if we want to make it zero for certain $k_y=k_{e}^{(l)}$, one just have to choose a relative phase: $\varphi_{l,e}=\pi-k_{e}^{(l)}$. In the case where there are more than two edge modes, one might be interested in cancelling the coupling to two or more momenta. A way of doing that would be to obtain an effective $G(k_y)$ out of the product required to cancel each momentum, separately. In the general case, where we are interested in suppressing $N_k$ modes, such proposal implies taking $G(k_y)$ as
\begin{equation}
G(k_y) = \prod_{\delta=0}^{N_k-1}G_{\varphi_\delta}(k_y)\;,
\end{equation}
where $\varphi_\delta=\pi-k_e^{(\delta)}$. With this choice, we get a non-local coupling in $\kk$-space of the form:
\begin{align}
G(k_y) &= \prod_{\delta=0}^{N_k-1}\left(1+e^{i(\varphi_\delta-k_y)}\right) \nonumber\\
&= \sum_{M=0}^{N_k-1}e^{-iMk_y}\sum_{|\Omega|=M}\exp\left(i\sum_{\delta\in\Omega}\varphi_\delta\right)\;,
\end{align}
where the sum over $|\Omega|=M$ denotes a sum over all the $N_k!/M!(N_k-M)!$ subsets of $\left\lbrace 1,2,...,N_k\right\rbrace$ with $M$ elements. From this expansion in powers of $e^{-ik_y}$, we can directly read the couplings in real-space:
\begin{equation}
g_{(0,y+M)}=\sum_{|\Omega|=M}\exp\left(i\sum_{\delta\in\Omega}\varphi_\delta\right)\;,
\end{equation}
with $M=0,...,N_k-1$. This shows that the number of non-local couplings required to cancel $N_k$ modes scales linearly with $N_k$, although this method will be eventually limited by the capacity to resolve the modes. As an example, for $N_k=2$, we have 
\begin{align}
   G_{\varphi_0,\varphi_1}(k_y)\propto 1+(e^{i\varphi_0}+e^{i\varphi_1})e^{ik_y}+e^{i(\varphi_0+\varphi_1)}e^{i2k_y}\,,
\end{align}
which can be achieved with real-space couplings of the form $g_{(0,y)}=g$, $g_{(0,y+1)}=ge^{i\varphi_0}+ge^{i\varphi_1}$ and $g_{(0,y+2)}=ge^{i(\varphi_0+\varphi_1)}$.\\

\bibliography{referencesAlex}
\end{document}